\documentclass[particles,article,submit,pdftex,oneauthor]{Definitions/mdpi} 
\firstpage{1} 
\pubvolume{1}
\issuenum{1}
\articlenumber{0}
\pubyear{2026}
\copyrightyear{2026}
\datereceived{ } 
\daterevised{ } 
\dateaccepted{ } 
\datepublished{ } 

\usepackage{slashed}
\usepackage{bm}
\usepackage{dsfont}

\preto{\abstractkeywords}{\nolinenumbers}

\newcommand{\conjg}[1]{\ensuremath{\hspace{1pt}\overline{\hspace{-1pt}#1\hspace{-1pt}}}\hspace{1pt}}
\newcommand{\vect}[1]{\bm{#1}}

\Title{Getting a handle on correlation functions}

\Author{Gernot Eichmann $^{1}$\orcidA{}}

\AuthorNames{Gernot Eichmann}

\address{%
$^{1}$ \quad Institute of Physics, University of Graz, NAWI Graz, Universitätsplatz 5, 8010 Graz, Austria; gernot.eichmann@uni-graz.at}

\abstract{
The central objects in a quantum field theory are its $n$-point correlation functions and matrix elements.
Their structure is determined by Lorentz invariance and leads to tensor decompositions  
whose Lorentz-invariant coefficient functions encode the physics of the process.
For growing $n$, the complexity of these objects may increase considerably
and make it challenging to deal with them.
Here we give a pedagogical introduction to the topic and provide some tools to manage
this complexity, and we will show how symmetries can be used as organizing principles.
}


\begin{document}


\section{Introduction}

\begin{figure}[t]
\includegraphics[width=1\textwidth]{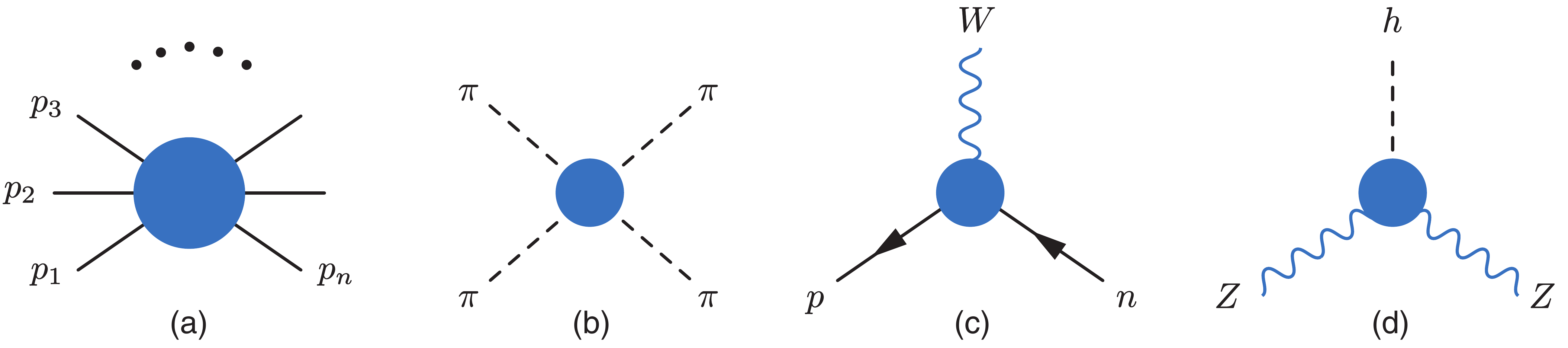}
\caption{(a) A generic $n$-point function has $n$ legs with $n$ momenta, of which only $n-1$ are independent due to momentum conservation. 
         Examples for $n$-point functions and matrix elements: (b) $\pi\pi$ scattering, (c) neutron $\beta$ decay, (d) Higgs-$Z$-boson coupling.\label{fig:npt-fct}}
\end{figure}

Pinning down the structure of $n$-point correlation functions and matrix elements is a common task  
in quantum field theory (QFT). They are the central objects of interest in many experiments as well as 
in perturbation theory, amplitude analyses, dispersion theory,  effective field theories, functional methods, lattice QFT and other approaches.
Loosely speaking, they describe everything that can happen when $n$ particles interact with each other, as pictured in Figure~\ref{fig:npt-fct}.
Typical examples in nuclear and particle physics are the scattering between electrons, pions or nucleons, weak decays like $\beta$ decay or  hadronic decays such as $\Sigma^+ \to \Lambda e^+ \nu_e$, three- and four-nucleon interactions, electromagnetic form factors, the anomalous magnetic moment of the muon, the three-gluon vertex, glueball amplitudes, the Higgs coupling to $Z$ bosons, or graviton amplitudes.

Formally, $n$-point correlation functions are the vacuum expectation values of time-ordered products of $n$ field operators at different coordinates, i.e.,
objects of the form
\begin{equation}\label{npt-fct-generic}
    \Gamma(x_1, \dots, x_n) = \langle 0 \, | \, \mathsf{T} \phi(x_1) \dots \phi(x_n) \,| \, 0 \rangle\,.
\end{equation}
They describe correlations between fields at different spacetime points.
A two-point function ($n=2$) is  called a propagator and represents the  amplitude for a particle to travel from point $x_1$ to $x_2$.
Correlation functions with $n\geq 3$  are also called vertices;
three-point functions ($n=3$) can describe creation and annihilation processes and higher $n$-point functions scattering processes. 

In the following we will use the name $n$-point function also for more general matrix elements,
where instead of the vacuum $|0\rangle$ the field operators in Eq.~\eqref{npt-fct-generic} can also be sandwiched  between onshell 
one-particle states $|p\rangle$ or multiparticle states $|p_1, p_2, \dots\rangle$. 
Matrix elements with the vacuum on one side and an onshell state on the other side
are called Bethe-Salpeter wave functions. 
Matrix elements without any field operators in the middle are  scattering amplitudes
and related to the residues of  $n$-point functions evaluated at the external propagator poles.

For the purposes of this work, we will usually not distinguish between these cases in our  terminology
and use the terms \textit{$n$-point correlation functions}, 
\textit{matrix elements} or  \textit{amplitudes} interchangeably. 
There is also no need to distinguish between full, connected and one-particle irreducible $n$-point functions.
The reason for this is that once we pass over to momentum space, all these objects have the same generic structure, namely
 \begin{equation}\label{nptfct-gen}
  \Gamma_{\alpha\beta\dots}^{\mu\nu\dots} (p_1, \dots p_n) = \sum_{i=1}^N f_i(p_1^2, p_2^2, p_1\cdot p_2, \dots)\,(\tau_i)_{\alpha\beta\dots}^{\mu\nu\dots} (p_1 \dots p_n) \,.
\end{equation}
The fields in Eq.~\eqref{npt-fct-generic} can belong to different representations of the Lorentz group and carry Dirac or Lorentz indices,
which are inherited by the amplitudes. As such, the amplitudes transform in a Lorentz-covariant way, and they  can be reconstructed 
from all possible Lorentz-covariant tensors transforming in the same way.
This leads to a maximum number $N$ of linearly independent  basis elements $(\tau_i)_{\alpha\beta\dots}^{\mu\nu\dots} (p_1 \dots p_n)$, 
which form a tensor basis for the amplitude.
Because all the Lorentz transformation properties are carried by the basis elements, the dressing functions $f_i(p_1^2, p_2^2, p_1\cdot p_2, \dots)$
must be Lorentz-invariant, so they can only depend on Lorentz-invariant momentum variables.
Ultimately those dressing functions are the main quantities of interest since they encode all the physics of the process.
The same statements  apply to onshell scattering amplitudes, because if one splits off the onshell spinors for fermion legs
(by replacing them with positive-energy projectors) or the polarization vectors for vector legs (by replacing them with transverse projectors),
the resulting Lorentz-covariant remainders have again the structure of Eq.~\eqref{nptfct-gen}. 

In the following, we will  only be interested in the \textit{structure} of $n$-point functions but not in their dynamics or how to calculate them.
From Eq.~\eqref{nptfct-gen} it is clear that higher $n$-point functions become progressively more complicated.
With more momenta in the system, the number of basis elements and  Lorentz invariants increases, which can complicate matters enormously. 
Especially for students and beginners, this can seem daunting. 
However, there are various techniques, some of them based on symmetries, which can serve as useful organizing principles to reduce this complexity.
Even though such tools are frequently employed, e.g. in functional methods~\cite{Eichmann:2011vu,Eichmann:2012mp,Eichmann:2014xya,Eichmann:2015nra,Eichmann:2016yit,Sanchis-Alepuz:2017jjd,Aguilar:2023qqd,Eichmann:2025etg,Braun:2025gvq},
they are often hidden in appendices or may seem overly technical.
To lower the entry bar for newcomers, 
we find it worthwhile to place them in the spotlight.  
The present work can be viewed as a technical companion to Refs.~\cite{Eichmann:2025wgs,Huber:2025cbd}, 
which are pedagogical introductions to functional methods, but as mentioned above,
the contents may also be useful to those in other research communities.

This work is organized as follows. In Section~\ref{sec:euc} we  motivate the Euclidean conventions that will be used throughout this work.
In Section~\ref{sec:2-2sc} we discuss $2\to 2$ scattering as an introductory example. 
In Section~\ref{sec:tensors} we investigate how many tensors there can be in general.
Section~\ref{sec:tensorbases} deals with the construction of tensor bases based on symmetries 
and Section~\ref{sec:li} with the choice of convenient momentum variables.
Finally, we conclude in Section~\ref{sec:conclusions}.

\begin{figure}[t]
\includegraphics[width=1\textwidth]{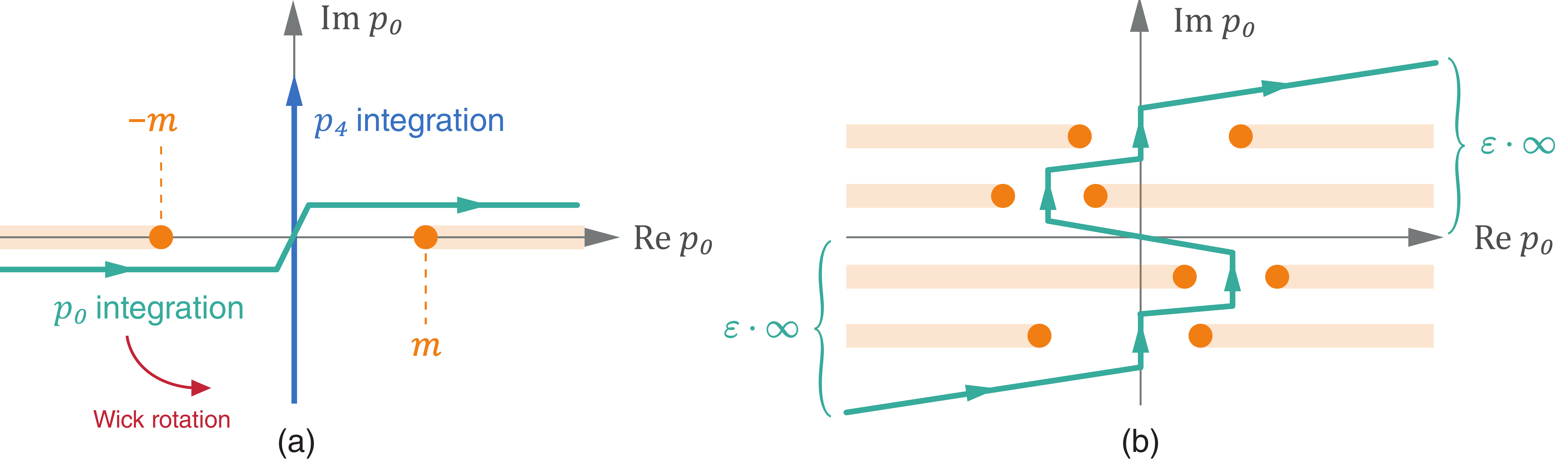}
\caption{(a) Loop integral with a simple pole. (b) Loop integral with multiple poles displaced from the real axis.
The orange tracks are the branch cuts arising from the $d^3p$ integration.\label{fig:wick}}
\end{figure}   
\unskip

\section{Preliminaries: Euclidean conventions}\label{sec:euc}

Throughout this work we will use Euclidean conventions, which amounts to replacing the Minkowski metric $g^{\mu\nu}$ 
with signature $(+,-,-,-)$ by a Euclidean metric $\delta^{\mu\nu}$ with signature $(+,+,+,+)$.
There are several good reasons for doing so, some of which are more practical and others more fundamental.
First of all, the Minkowski metric is not very convenient. 
Fighting with upper and lower indices can become  painful if the expressions are long;
different authors use different sign conventions for the Levi-Civita symbol $\varepsilon^{\mu\nu\rho\sigma}$, etc.
In Euclidean conventions, upper and lower indices are the same. 
This  makes it easy to write code, since you never need to worry about squeezing in metric tensors:
a Lorentz tensor $A^{\mu\nu}$ is simply a $4\times 4$ matrix, and the product of two Lorentz tensors $A^{\mu\alpha} A^{\alpha\nu}$
is a matrix multiplication.

Another reason is that we often deal with spacelike momenta, i.e., virtual particles.
In particular, loop momenta are spacelike. Sometimes it is emphasized in a paper that the work is done  
in Minkowski space, but eventually  a Wick rotation is performed to compute loop integrals --
but from that point onwards it is   Euclidean.
So why not start directly with a Euclidean metric?
In fact, one can make the point that QFT  is already `Euclidean':
to make the partition function well-defined one needs imaginary-time boundary conditions, so
every integral over $d^4x$ or $d^4p$ really means
\begin{equation}
   \int d^4x = \int d^3x \int_{-\infty(1-i\varepsilon)}^{\infty(1-i\varepsilon)}dx_0 \quad \Leftrightarrow \quad
   \int d^4p = \int d^3p \int_{-\infty(1+i\varepsilon)}^{\infty(1+i\varepsilon)}dp_0\,.
\end{equation}
This is usually implemented through the $i\varepsilon$ prescription
by writing $p_0^2 = \vect{p}^2 + m^2 -i\varepsilon$, 
which for an integral with a propagator pole entails
\begin{equation}
    \int d^3p \int_{-\infty}^\infty dp_0 \,\frac{1}{p^2 - m^2 +i\varepsilon} \,(\dots)\,.
\end{equation}
The propagator has poles at $p_0 = \pm \sqrt{\vect{p}^2 + m^2}$,
which move along the real $p_0$ axis depending on the value of $\vect{p}^2$  (see Figure~\ref{fig:wick}a). If we exchange the order of integrations,
these poles turn into branch cuts  starting at $p_0 = \pm m$ which extend to infinity.
According to the $i\varepsilon$ prescription we should then start the $p_0$ integration slightly below the real axis
and stop slightly above. But as long as the integrand falls off fast enough at complex infinity, we can perform a Wick rotation 
and integrate along the imaginary $p_0$ axis, which is the real $p_4 = ip_0$ axis. Therefore, had we  started directly
in Euclidean space, we could just have ploughed straight along the real $p_4$ axis without encountering any cut.

Things can become more complicated if more propagator poles are involved.
Sometimes we would like to calculate  integrals for complex external momenta. For example, 
suppose an amplitude  depends on the photon virtuality $Q^2$ and we would like to compute
it for complex $Q^2$. In that case, $Q^2$ also enters in the propagators in the loop 
and the cuts usually align as sketched in Figure~\ref{fig:wick}b, with
a finite distance from the real axis. The $i\varepsilon$ in the propagators then becomes meaningless,
but the boundary conditions still work out: We should start the integration
on the bottom left and end up  top right, without crossing any cuts.
The result of the integration is a Lorentz-invariant, analytic function in $Q^2$,
but this can only work as long as we do not cross any  branch cut along the way.
In this sense, there is actually no difference between Minkowski and Euclidean because once
the cuts cross the imaginary axis,  the  contour needs to be deformed either way.
Since this usually happens for (large) timelike  momenta,  the statement `\textit{We need to go to Minkowski space}'
really just means `\textit{We need to go to timelike momenta}', which requires contour deformations or
residue calculus like in Figure~\ref{fig:wick}b, or analytic continuations or some other method. 
That being said, we will never actually integrate anything in the following, 
so the discussion is rather academic in our context. Still, it serves as a motivation why a Euclidean metric
is  as good as a Minkowski one, or even simpler for the reasons stated earlier.

To make things concrete, a Euclidean four-vector is given by
\begin{equation}
   a^\mu_E = \left[ \vect{a} \atop ia_0 \right].
\end{equation}
This implies that scalar products switch signs: $a_E\cdot b_E = -a\cdot b$ and $a_E^2 = -a^2$.
To preserve the meaning of a Feynman slash $\slashed{a} = a^0 \gamma^0 - \vect{a}\cdot\vect{\gamma}$,
we must also redefine the $\gamma$ matrices:
\begin{equation}
   i\gamma^\mu_E = \left[ \vect{\gamma} \atop i\gamma_0 \right], \quad \gamma^5_E = \gamma^5 \qquad \Rightarrow \qquad \slashed{a}_E = a_E \cdot \gamma_E = i\slashed{a}\,.
\end{equation}
The Clifford algebra relation $\{\gamma^\mu_E, \gamma^\nu_E \} = 2\delta^{\mu\nu}$ then implies $(\gamma^i_E)^2 = 1$ and $\gamma^\mu_E = (\gamma^\mu_E)^\dag$.
Denoting the Pauli matrices by $\sigma_k$, the Euclidean $\gamma$ matrices in the standard representation read
\begin{equation}
   \gamma_E^k = \left( \begin{array}{cc} 0 & -i\sigma_k \\ i\sigma_k & 0 \end{array}\right), \qquad
   \gamma_E^4 = \left( \begin{array}{cc} \mathds{1} & 0 \\ 0 & -\mathds{1} \end{array}\right), \qquad
   \gamma_E^5 = \left( \begin{array}{cc} 0 & \mathds{1} \\ \mathds{1} & 0 \end{array}\right).
\end{equation}

From now on we  drop the subscript $E$ and all expressions will be Euclidean.  
A momentum $p$ is spacelike if $p^2 > 0$ and timelike if $p^2 < 0$.
We will rarely need to go into a specific frame since all relations can be written in a Lorentz-covariant or even Lorentz-invariant way.
Still, for an onshell momentum in the rest frame we would write
\begin{equation}\label{restframe}
    p = \left[ \begin{array}{c} 0 \\ 0 \\ 0 \\ im \end{array}\right]  \quad \Rightarrow \quad p^2 = -m^2\,.
\end{equation}
A general four-momentum (like a loop momentum) can be expressed in hyperspherical variables,
\begin{equation}
    p = \sqrt{p^2} \left[ \begin{array}{l} \sqrt{1-z^2}\sqrt{1-y^2}\,\sin\psi \\ \sqrt{1-z^2}\sqrt{1-y^2}\,\cos\psi \\ \sqrt{1-z^2}\,y \\ z \end{array}\right],
\end{equation}
and a four-momentum integration reads
\begin{equation}
   \int \!\frac{d^4p}{(2\pi)^4} = \frac{1}{(2\pi)^4} \,\frac{1}{2} \int_0^\infty dp^2\,p^2 \int_{-1}^1 dz \,\sqrt{1-z^2} \int_{-1}^1 dy \int_0^{2\pi} d\psi\,.
\end{equation}
More relations and a dictionary between Minkowski and Euclidean conventions can be found, e.g., in the appendices of Refs.~\cite{Eichmann:2016yit,Eichmann:2018ytt}.

\section{Example: Two-particle scattering}\label{sec:2-2sc}

\begin{figure}[t]
\begin{adjustwidth}{-\extralength}{0cm}
\includegraphics[width=18.5cm]{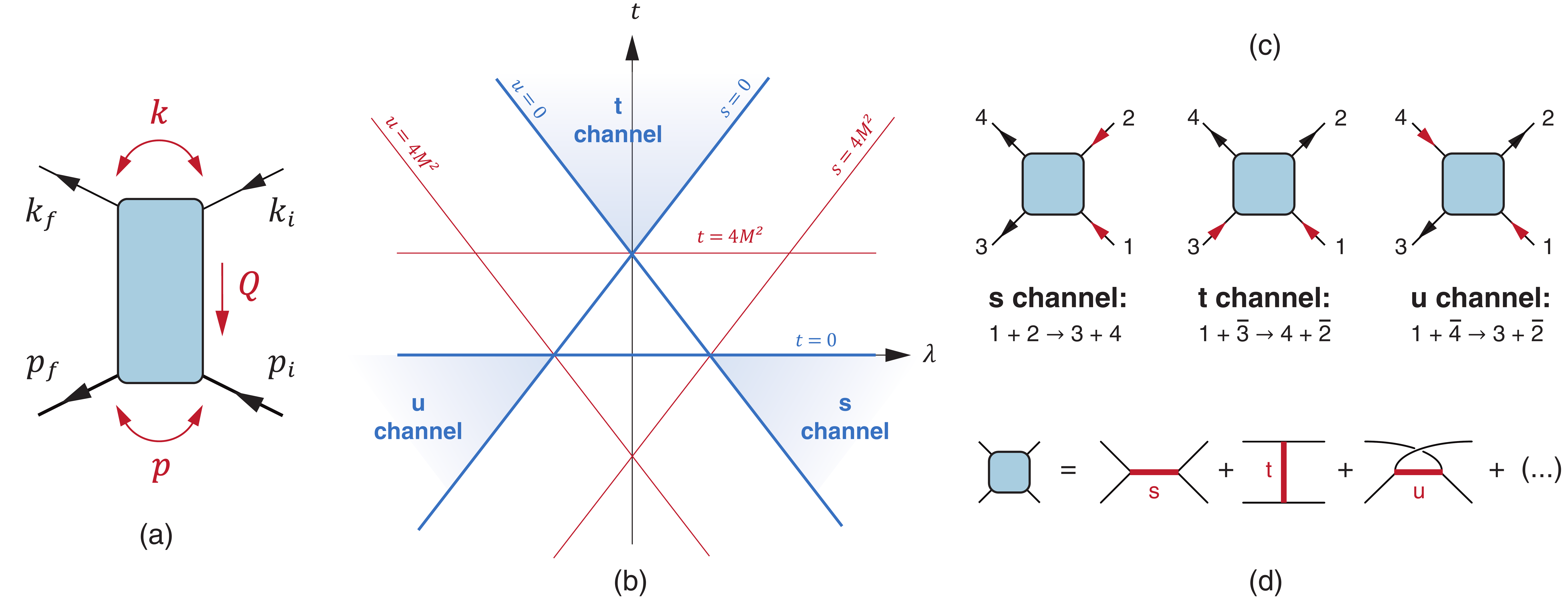}
\caption{(a) Kinematics in the $2\to 2$ scattering amplitude. (b) Mandelstam plane in the variables $\lambda$ and $t$.
         (c) The different $s$-, $t$- and $u$-channel processes are described by the same amplitude. (d) The $s$-, $t$- and $u$-channel poles
         are responsible for the main momentum dependence of the amplitude. \label{fig:mandelstam}}
\end{adjustwidth}
\end{figure}

Let us begin with a standard textbook example, namely the scattering of two scalar particles  sketched in Figure~\ref{fig:mandelstam}a.  
This is the simplest example of a four-point function.
Because the particles are scalar, there is only one Lorentz-invariant function $\Gamma(p_i, k_i, p_f, k_f)$.
Momentum conservation entails that only three  momenta are independent. It is useful to work with the average momenta $p$ and $k$ 
and the momentum transfer $Q$,
\begin{equation}
    p = \frac{p_i + p_f}{2}\,, \qquad k = \frac{k_i + k_f}{2}\,, \qquad Q = p_f - p_i = k_i - k_f\,,
\end{equation}
and therefore
\begin{equation}
   p_i = p - \frac{Q}{2}\,, \qquad 
   p_f = p + \frac{Q}{2}\,, \qquad
   k_i = k + \frac{Q}{2}\,, \qquad
   k_f = k - \frac{Q}{2}\,.
\end{equation}

From three momenta one can form six Lorentz invariants: $p^2$, $k^2$, $Q^2$, $p\cdot Q$, $k\cdot Q$, and $p\cdot k$.
If all particles are onshell, i.e., $p_i^2 = p_f^2 = -M^2$ and $k_i^2 = k_f^2 = -m^2$, then
\begin{equation}
   \begin{array}{rl}
      p_i^2 &\!\!\!= p^2 + \displaystyle\frac{Q^2}{4} - p\cdot Q = -M^2\,,  \\[4mm]
      p_f^2 &\!\!\!= p^2 + \displaystyle\frac{Q^2}{4} + p\cdot Q = -M^2\,, 
   \end{array}\qquad
   \begin{array}{rl}
      k_i^2 &\!\!\!= k^2 + \displaystyle\frac{Q^2}{4} + k\cdot Q = -m^2\,, \\[4mm]
      k_f^2 &\!\!\!= k^2 - \displaystyle\frac{Q^2}{4} - k\cdot Q = -m^2\,, 
   \end{array}
\end{equation}
and therefore
\begin{equation}
   p\cdot Q = k\cdot Q = 0\,, \qquad p^2 = -M^2 - \frac{Q^2}{4}\,, \qquad k^2 = -m^2 - \frac{Q^2}{4}\,.
\end{equation}
This leaves only two independent variables, namely $Q^2$ and $p\cdot k$.
Their dimensionless versions define the momentum transfer $\tau$ and the crossing variable $\lambda$:
\begin{equation}
    \tau = \frac{Q^2}{4M^2}\,, \qquad \lambda = -\frac{p\cdot k}{M^2}\,.
\end{equation}
Let us also define the  constant
\begin{equation}
   \varepsilon = \frac{M^2-m^2}{2M^2}  \quad \Rightarrow \quad m^2 = M^2 \,(1-2\varepsilon)\,, \quad
   p^2 + k^2 = -2M^2 (1+\tau-\varepsilon)\,.
\end{equation}
Alternatively, one can define the Mandelstam variables $s$, $u$ and $t$:
\begin{equation*}
\begin{split}
   s &= -(p_i + k_i)^2 = -(p_f + k_f)^2 = -(p+k)^2 = -(p^2 + k^2+2p\cdot k) = 2M^2 \,(1+\tau + \lambda - \varepsilon)\,, \\ 
   u &= -(p_i - k_f)^2 = -(p_f - k_i)^2 = -(p-k)^2 = -(p^2 + k^2-2p\cdot k) = 2M^2 \,(1+\tau - \lambda - \varepsilon)\,, \\ 
   t &= -(p_f - p_i)^2 = -(k_i - k_f)^2 = -Q^2 = -4M^2 \tau\,.
\end{split}
\end{equation*}
They are not independent because their sum $s+t+u = 4M^2 (1-\varepsilon) = 2M^2 + 2m^2$ is the sum of the squared masses. 
Here you also see how the cumbersome minus sign between Minkowski and Euclidean scalar products is handled in practice:
One simply defines new  variables which mean the same in both conventions.
The amplitude $\Gamma(p_i, k_i, p_f, k_f) = \Gamma(s,t,u) = \Gamma(\tau,\lambda)$ is then a Lorentz-invariant function
of two Lorentz-invariant  variables, which do not know anything about `Minkowski' or `Euclidean'.
The name \textit{crossing variable} comes from the fact that $\lambda$ is odd under $s-u$ exchange:
\begin{equation}
  \tau = -\frac{t}{4M^2} = \frac{s+u}{4M^2} - 1 + \varepsilon\,, \qquad \lambda = \frac{s-u}{4M^2}\,.
\end{equation}

\unskip 
 
 The kinematic domain of $\Gamma(\tau,\lambda)$ can be visualized by the Mandelstam plane in Figure~\ref{fig:mandelstam}b,
where it is shown for the special case of identical masses ($\varepsilon=0$).
The same amplitude describes three different processes with different physical domains (Figure~\ref{fig:mandelstam}c).
The $s-$channel reaction is the process $1 + 2 \to 3 + 4$, the $t-$channel process corresponds to $1 + \conjg{3} \to 4 + \conjg{2}$
and the $u-$channel to $1 + \conjg{4} \to 3 + \conjg{2}$.
The momentum dependence of the amplitude is  governed by its singularities, which are the physical poles and cuts
where intermediate particles can go onshell (Figure~\ref{fig:mandelstam}d).
For example in $\pi\pi$ scattering, two pions can create further (e.g., vector) mesons with masses $m_i$ above the two-pion threshold, which 
produce poles at $s = m_i^2$, $t = m_i^2$ and $u = m_i^2$. 
Since they line up symmetrically in $s$, $t$ and $u$, the Mandelstam plane has a symmetry under rotation by 120$^\circ$.
This already tells us  a great deal about the momentum dependence of the amplitude $\Gamma(\tau,\lambda)$!

Another point is that  $\Gamma(\tau,\lambda)$ is Lorentz-invariant, so there is no need to go into a specific frame unless to relate with experiments.
For example in the center-of-mass frame, where $|\vect{p}|\,\vect{e}$ and $|\vect{p}|\,\vect{e}'$ are 
the three-momenta corresponding to $p_i$ and $p_f$, respectively, and $\vect{e}\cdot\vect{e}' = \cos\theta_\text{CM}$
defines the scattering angle, we have for $\varepsilon = 0$: 
\begin{equation}
\begin{split}
    \tau = \frac{\vect{p^2}}{2M^2} \,\left( 1 - \cos\theta_\text{CM} \right), \qquad
    \lambda = 1 + \frac{\vect{p}^2}{2M^2}\,(3+\cos\theta_\text{CM})\,.
\end{split}
\end{equation}
In a different frame, $\tau$ and $\lambda$ are related differently to the three-momentum and scattering angle,
but since $\tau$ and $\lambda$ are Lorentz-invariant they are the same in any frame.

Returning to Eq.~\eqref{nptfct-gen}, in general the situation can be much more complicated
than the simple example above. 
$n$-point functions usually have a Dirac and/or Lorentz structure, so there is not just one
dressing function but there can be many. Furthermore, the amplitudes are usually offshell,
so there can be more (or many) Lorentz invariants -- in our example we would need to go back to the original six instead of just $\tau$ and $\lambda$.
These issues complicate the problem both algebraically and numerically.
How many tensors are there? What are the most convenient choices for the tensor bases? 
What is the best way to organize the Lorentz invariants?
Are the dressing functions still going to fit into the computer's memory?
We will discuss these points in the following sections.
Along the way we will see that permutation symmetries, but also gauge symmetries, are very useful in this regard
and  can simplify matters a lot.

\section{How many tensors are there?} \label{sec:tensors}

Before we talk about symmetries, let us return to Eq.~\eqref{nptfct-gen} and   ask: How many tensors are there, i.e., what is $N$?
Clearly, this number depends on the process. But for a given process, what are the rules to determine $N$?
As mentioned in the introduction, the contents of this section are scattered over the appendices of various papers and
we will draw extensively from Refs.~\cite{Eichmann:2011vu,Eichmann:2012mp,Eichmann:2015nra,Eichmann:2016yit,Wallbott:2019dng}.

\begin{figure}[t]
\includegraphics[width=1\textwidth]{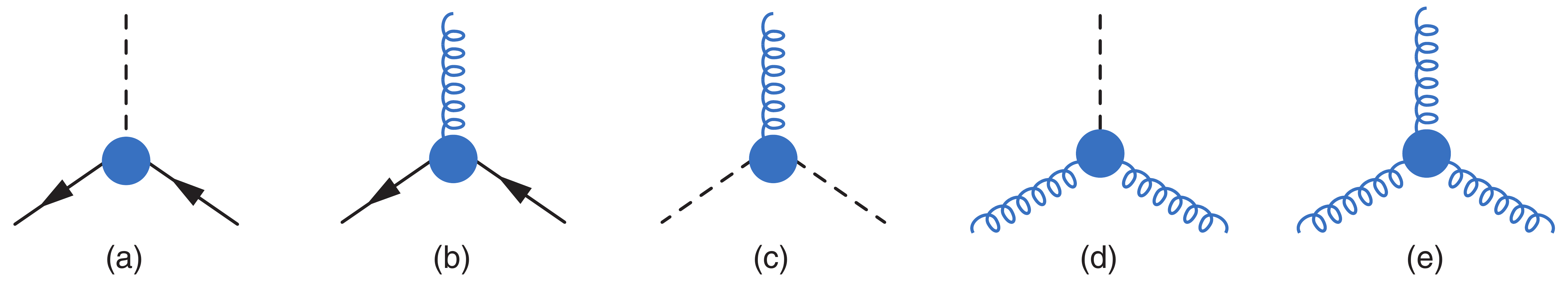}
\caption{Examples of three-point functions. Solid lines represent fermions, dashed lines  scalars or pseudoscalars,
and springs are vectors or axialvectors.\label{fig:3pt-fct}}
\end{figure}

As an example, consider a fermion-scalar vertex like in Figure~\ref{fig:3pt-fct}a,
i.e., a three-point function with a fermion, an antifermion and a scalar leg. 
This object describes the coupling of a scalar particle to a fermion and could represent, e.g., 
a scalar bound state made of a quark and an antiquark. 
It depends on two independent momenta, which we call  $p$ and $q$; it does not matter
which particles they actually represent.
The vertex has a fermion and an antifermion leg, so it must have two Dirac indices, i.e.,  it is a matrix in Dirac space. 
From $1$, $\gamma_5$, $\gamma^\mu$, $\gamma^\mu \gamma_5$ and $[\gamma^\mu,\gamma^\nu]$,
the only possible Lorentz-covariant objects that are compatible with such a structure are $1$, $\slashed{p}$, $\slashed{q}$ and $[\slashed{p},\,\slashed{q}]$.
We cannot use $\gamma_5$ since this would return the wrong parity, and $[\slashed{p},\slashed{p}]=0$.
Thus, the general structure of the vertex is
\begin{equation}
  \Gamma_{\alpha\beta} (p,q) = \sum_{i=1}^4 f_i(p^2, q^2, p\cdot q)\,(\tau_i)_{\alpha\beta} (p,q)\,, \qquad
  \tau_i(p,q) \in \{ 1\,, \; \slashed{p}\,, \; \slashed{q}\,, \; [\slashed{p},\,\slashed{q}] \}\,.
\end{equation}

What is the number of basis tensors for a general $n$-point function?
Let us suppress the Lorentz and Dirac indices and write the $n$-point function as $\Gamma(p, q, k, l, \dots)$, 
where $p$, $q$, $k$, $l$, \dots are the $n-1$ independent momenta.
We can always choose a reference frame such that the first vector points in the  fourth direction, the second in the third and fourth directions, and so on:
\begin{equation}\label{specific-frame}
   p = \left[ \begin{array}{c} 0 \\ 0 \\ 0 \\ \bullet \end{array}\right], \quad
   q = \left[ \begin{array}{c} 0 \\ 0 \\ \bullet \\ \bullet \end{array}\right], \quad
   k = \left[ \begin{array}{c} 0 \\ \bullet \\ \bullet \\ \bullet \end{array}\right], \quad
   l = \left[ \begin{array}{c} \bullet \\ \bullet \\ \bullet \\ \bullet \end{array}\right], \quad \dots
\end{equation}
The circles refer to non-zero entries. 
From these momenta we can construct orthonormal unit vectors using Gram-Schmidt orthogonalization:
\begin{equation}\label{specific-frame-2}
\begin{split}
   n_4 &= \text{Hat}\,[p] = \left[ \begin{array}{c} 0 \\ 0 \\ 0 \\ 1 \end{array}\right],  \quad
   n_3 = \text{Hat}\,\left[ q-(q\cdot n_4) \,n_4 \right] = \left[ \begin{array}{c} 0 \\ 0 \\ 1 \\ 0 \end{array}\right], \\
   n_2 &= \text{Hat}\left[ k- \sum_{i=3}^4 (k\cdot n_i) \,n_i \right] = \left[ \begin{array}{c} 0 \\ 1 \\ 0 \\ 0 \end{array}\right],  \quad
   n_1 = \text{Hat}\left[ l- \sum_{i=2}^4 (l\cdot n_i) \,n_i \right] = \left[ \begin{array}{c} 1 \\ 0 \\ 0 \\ 0 \end{array}\right].
\end{split}
\end{equation}
The `Hat'  means normalization, i.e., $\text{Hat}\,[p] = p / \sqrt{p^2}$.
Note that $\sqrt{p^2}$ can also be complex, like in the example of the rest-frame vector from Eq.~\eqref{restframe} with $\sqrt{p^2} = im$.

The definitions of the $n_i$ above are Lorentz-covariant. If we use a frame like in Eq.~\eqref{specific-frame},
the $n_i$ are the Euclidean unit vectors as given above, otherwise they  become more complicated.
In any case, note that the orthogonalization stops at $n_1$ because we are limited by four spacetime dimensions.
Therefore, irrespective of how many momenta appear in the $n$-point function, 
in the basis construction there are at most four momenta available as building blocks!

\unskip

\subsection{Three-point functions}

To exemplify this, consider the three-point functions in Figure~\ref{fig:3pt-fct}.
They depend on two independent momenta, 
so for the basis construction we only have the unit vectors $n_3$ and $n_4$
at our disposal. For the fermion-scalar vertex (Figure~\ref{fig:3pt-fct}a)  this yields the basis
\begin{equation}\label{tau_i-sc}
   \left\{  1\,, \; \slashed{n}_3\,, \; \slashed{n}_4\,, \; \slashed{n}_3\,\slashed{n}_4 \right\} .
\end{equation}
Because $\slashed{n}_3^2 = \slashed{n}_4^2 = 1$ and $\slashed{n}_3\,\slashed{n}_4 = -\slashed{n}_4\,\slashed{n}_3$, these are all possible options.
For a fermion-vector vertex (Figure~\ref{fig:3pt-fct}b) there are twelve possible combinations:
\begin{equation} \label{3pt:fv-vertex}
     \left\{ \gamma^\mu, \; n_3^\mu, \; n_4^\mu \right\} \times \left\{  1\,, \; \slashed{n}_3\,, \; \slashed{n}_4\,, \; \slashed{n}_3\,\slashed{n}_4 \right\} .
\end{equation}
A scalar-vector vertex (Figure~\ref{fig:3pt-fct}c) has only two tensors,
\begin{equation}\label{3pt:sv-vertex}
    n_3^\mu\,, \quad n_4^\mu \,,
\end{equation}
and a scalar-two-vector vertex (Figure~\ref{fig:3pt-fct}d) like a scalar-two-photon vertex or a scalar glueball amplitude  has five:
\begin{equation}\label{3pt:svv-vertex}
    \underbrace{\delta^{\mu\nu}}_{1}\,, \quad \underbrace{n_i^\mu\,n_j^\nu}_{2^2=4} \,.
\end{equation}
For an amplitude made of three vector particles, like the three-gluon vertex (Figure~\ref{fig:3pt-fct}e), there are 14:
\begin{equation}\label{3pt:3gv}
    \underbrace{ n_i^\mu\,n_j^\nu\,n_k^\rho}_{2^3=8}\,, \quad
    \underbrace{\delta^{\mu\nu}\,n_i^\rho}_{2}\,, \quad
    \underbrace{\delta^{\nu\rho}\,n_i^\mu}_{2}\,, \quad
    \underbrace{\delta^{\rho\mu}\,n_i^\nu}_{2}\,.
\end{equation}

What happens in the case of opposite parity? As long as fermions are involved, the answer is simple
because one just needs to attach a factor $\gamma_5$. For example, 
the basis for a fermion-pseudoscalar vertex (like the Bethe-Salpeter amplitude of a pion) 
is identical to Eq.~\eqref{tau_i-sc} except for an additional $\gamma_5$:
\begin{equation}
   \gamma_5 \left\{  1\,, \; \slashed{n}_3\,, \; \slashed{n}_4\,, \; \slashed{n}_3\,\slashed{n}_4 \right\}.
\end{equation}
The same applies to a fermion-axialvector vertex, which still has 12 tensors:
\begin{equation} \label{3pt:fav-vertex}
     \gamma_5 \times \left\{ \gamma^\mu, \; n_3^\mu, \; n_4^\mu \right\} \times \left\{  1\,, \; \slashed{n}_3\,, \; \slashed{n}_4\,, \; \slashed{n}_3\,\slashed{n}_4 \right\} .
\end{equation}
In the absence of fermions, one has to work with the antisymmetric Levi-Civita symbol $\varepsilon^{\mu\nu\alpha\beta}$,
with $\varepsilon^{1234} = 1$ in Euclidean conventions (remember that there is no distinction between upper and lower indices).
From two momenta $n_3$ and $n_4$  we can only construct the combinations
\begin{equation}\label{epsilon-tensors}
   \varepsilon^{\mu\nu} = \varepsilon^{\mu\nu\alpha\beta} n_3^\alpha\,n_4^\beta\,, \qquad
   \varepsilon^{\mu\nu\rho\alpha} n_3^\alpha\,, \qquad \varepsilon^{\mu\nu\rho\alpha} n_4^\alpha\,.
\end{equation}
This entails that a scalar-axialvector vertex like in Figure~\ref{fig:3pt-fct}c does not exist,
because one cannot construct an axialvector with one Lorentz index from two momenta only.
For the same reason, odd-parity three-point functions with three pseudoscalars or one pseudoscalar and two scalars
also  do not exist.
For a pseudoscalar-two-vector vertex (Figure~\ref{fig:3pt-fct}d) like the $\pi\gamma\gamma$ amplitude  
the only possible tensor is $\varepsilon^{\mu\nu}$ in Eq.~\eqref{epsilon-tensors}. For a three-point function with three axialvectors,
or one axialvector and two vectors, there are six basis elements, namely the two elements $\varepsilon^{\mu\nu} n_i^\rho$
plus their permutations. The two  additional tensors $\varepsilon^{\mu\nu\rho\alpha} n_i^\alpha$ from Eq.~\eqref{epsilon-tensors}
linearly depend on those due to the Schouten identity 
\begin{equation}
   \varepsilon^{\mu\nu\rho\sigma} p^\tau + \varepsilon^{\nu\rho\sigma\tau} p^\mu + \varepsilon^{\rho\sigma\tau\mu} p^\nu + \varepsilon^{\sigma\tau\mu\nu} p^\rho + \varepsilon^{\tau\mu\nu\rho} p^\sigma = 0\,,
\end{equation}
which holds for any four-vector $p$   
and reflects the fact that an antisymmetrization over more indices than there are  dimensions in spacetime must return zero.

Tensor bases that are constructed from orthogonal unit vectors are  easy to orthogonalize.
If we call the tensors in Eq.~\eqref{tau_i-sc} $\tau_i$ and define  `conjugate' tensors $\conjg{\tau}_i$ by 
\begin{equation}
   \left\{  1\,, \; \slashed{n}_3\,, \; \slashed{n}_4\,, \; \slashed{n}_4\,\slashed{n}_3 \right\} ,
\end{equation}
this gives the orthonormality relation
\begin{equation}
    \frac{1}{4}\,\text{Tr}\left\{ \conjg{\tau}_i\,\tau_j \right\} = \delta_{ij}\,.
\end{equation}
Orthogonal tensor bases are extremely useful in the solution of dynamical equations.
Given some equation of the form $\Gamma = A(\Gamma)$, where $A$ can be a linear or non-linear integral or differential operator,
applying the orthonormality relation projects out the dressing functions:
\begin{equation}
    \Gamma = \sum_i f_i\,\tau_i = A(\Gamma) \quad \Rightarrow \quad f_i 
    = \frac{1}{4}\,\text{Tr}\left\{ \conjg{\tau}_i\,A(\Gamma)\right\}.
\end{equation}
Since the right-hand side depends on (integrals or derivatives of) the $f_j$,  one immediately arrives at Lorentz-invariant equations
which couple the $f_i$ together.
Or if the right-hand side does not depend on $\Gamma$, it still tells you how to project out the dressing functions.

\subsection{Four-point functions}

\begin{figure}[t]
\includegraphics[width=1\textwidth]{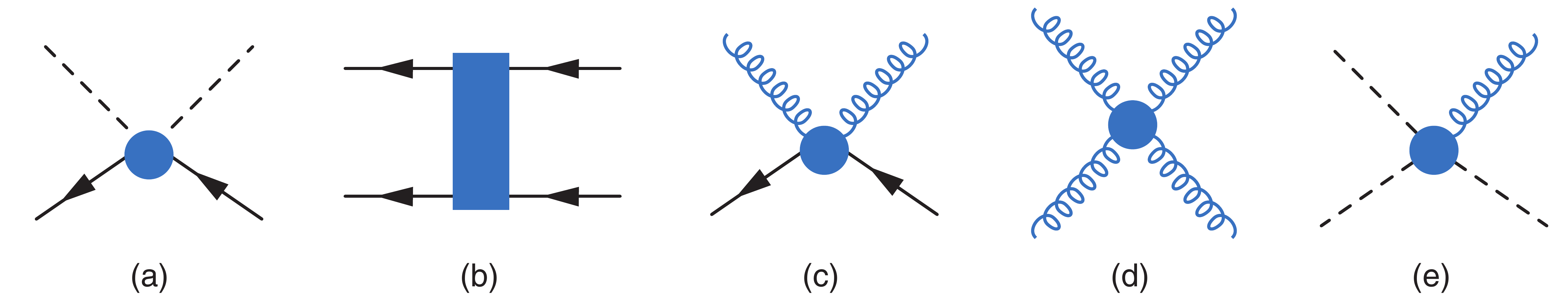}
\caption{Examples of four-point functions. Solid lines represent fermions, dashed lines  scalars or pseudoscalars,
and springs are vectors or axialvectors.
\label{fig:4pt-fct}}
\vspace{5mm}
\end{figure}

What about four-point functions, like those in Figure~\ref{fig:4pt-fct}?
A four-point function depends on three momenta, so we can choose $n_2$, $n_3$, $n_4$ as unit vectors.
But with three vectors we can also do the following:
\begin{equation}
   v^\mu = \varepsilon^{\mu\alpha\beta\gamma}\,n_2^\alpha\,n_3^\beta\,n_4^\gamma = \left[ \begin{array}{c} 1 \\ 0 \\ 0 \\ 0 \end{array}\right].
\end{equation}
This definition is again Lorentz-covariant, while  the last equality only holds in the specific frame from Eq.~\eqref{specific-frame} 
where $v^\mu$ becomes the unit vector in direction 1.
The Levi-Civita symbol changes parity, so $v^\mu$ is  an axialvector and we will need an even number of $v$'s to
construct a tensor basis with positive parity. In any case, now we can write 
\begin{equation}\label{v-relations}
\begin{split}
   \delta^{\mu\nu} = v^\mu \,v^\nu + \sum_{i=1}^3 n_i^\mu\,n_i^\nu\,,  \qquad 
   \gamma^\mu = v^\mu \slashed{v} + \sum_{i=1}^3 n_i^\mu\,\slashed{n}_i\,, \qquad
   \slashed{v} = \gamma_5\,\slashed{n}_2\,\slashed{n}_3\,\slashed{n}_4\,.
\end{split}
\end{equation}
These relations are obviously true in the simple frame~\eqref{specific-frame-2}, where $v^\mu$ and the $n_i^\mu$ 
are the Euclidean unit vectors (and furthermore $\slashed{v} = \gamma_1$ and $\slashed{n}_i = \gamma_i$). 
However, because these relations are Lorentz-covariant  they must hold in any frame.
This means we no longer need $\delta^{\mu\nu}$ and $\gamma^\mu$ in the basis construction,
the $n_i^\mu$ and $v^\mu$ are enough!

This observation simplifies the basis construction considerably.
As an example, consider a fermion-two-scalar vertex (Figure~\ref{fig:4pt-fct}a). It depends on eight possible tensors:
\begin{equation}\label{f2s-vertex}
   \Omega = \big\{ \underbrace{1}_{1}\,, \; \underbrace{\slashed{n}_i}_{3}\,, \; \underbrace{\slashed{n}_i\,\slashed{n}_j}_{3}\,, \; \underbrace{\slashed{n}_i\,\slashed{n}_j\,\slashed{n}_k}_{1} \big\}, \qquad i < j < k\,.
\end{equation}
Note that $\slashed{n}_2\,\slashed{n}_3\,\slashed{n}_4 = \gamma_5\,\slashed{v}$ from Eq.~\eqref{v-relations},
so we could equally write the basis  as
\begin{equation}
   \underbrace{\big\{ 1\,, \; \gamma_5\,\slashed{v} \big\}}_{2} \times \underbrace{\big\{ 1\,, \; \slashed{n}_i  \big\}}_{4}\,.
\end{equation}

\unskip

What about a fermion four-point function (Figure~\ref{fig:4pt-fct}b)? In that case there are 128 tensors:
\begin{equation}\label{f-4pt-fct}
   \underbrace{\Omega_{\alpha\beta}\,\Omega_{\gamma\delta}}_{8^2 = 64}\,, \qquad \underbrace{(\gamma_5 \,\Omega)_{\alpha\beta}\,(\gamma_5 \, \Omega)_{\gamma\delta}}_{8^2 = 64}\,.
\end{equation}
Any other combination like $\gamma^\mu \,\Omega \otimes \gamma^\mu \, \Omega$ is linearly dependent on those,
because we can express $\gamma^\mu$ purely in terms of slashes of  $n_i^\mu$ and $v^\mu$ through Eq.~\eqref{v-relations}!
For a quark-antiquark four-point function there are in addition two color tensors, which makes $128 \times 2 = 256$ Dirac-color tensors in total.
(This might be a good point to mention that Eq.~\eqref{nptfct-gen} can generally have a flavor and color structure as well, 
which is however not the focus of the present work -- here we only consider the Dirac and Lorentz structure.)

Another example is a fermion-two-vector vertex (or fermion Compton vertex) like in Figure~\ref{fig:4pt-fct}c, in which case there are also 128 tensors~\cite{Eichmann:2012mp}:
\begin{equation}\label{4pt:fcompton}
   \big\{ \underbrace{v^\mu v^\nu}_{1} \,, \; \underbrace{n_i^\mu\,n_j^\nu}_{3^2=9} \big\} \times \underbrace{\Omega}_{8}\,, \qquad 
   \big\{ \underbrace{v^\mu n_i^\nu}_{3}\,, \; \underbrace{n_i^\mu\,v^\nu}_{3} \big\} \times \underbrace{\gamma_5\,\Omega}_{8}\,.
\end{equation}
Note that $v^\mu$ and $\gamma_5$ must come in pairs to preserve the correct parity.
Similarly, an amplitude with four vector legs, like the four-photon vertex (light-by-light amplitude) or four-gluon vertex  
in Figure~\ref{fig:4pt-fct}d,
depends on 136 tensors~\cite{Eichmann:2014ooa,Eichmann:2015nra}:
\begin{equation}\label{4pt:lbl}
    \underbrace{v^\mu \,v^\nu \,v^\rho \,v^\sigma}_{1}\,, \qquad 
    \underbrace{v^\mu \,v^\nu \,n_i^\rho \,n_j^\sigma}_{3^2=9, \atop 6 \,\text{permutations}}\,, \qquad 
    \underbrace{n_i^\mu \,n_j^\nu \,n_k^\rho \,n_l^\sigma}_{3^4 = 81}\,.
\end{equation}

The case with only one vector leg and three scalar legs (Figure~\ref{fig:4pt-fct}e) 
is simple, as it only allows for the three tensors $n_i^\mu$.

\newpage

The construction of four-point functions with opposite parity is particularly simple. If fermions are involved, one only needs to multiply with $\gamma_5$.
For instance in Eq.~\eqref{4pt:fcompton} for a fermion-vector-axialvector vertex, this swaps the position of  $\gamma_5$.
For a vertex like in Figure~\ref{fig:4pt-fct}d with one or three axialvector legs, the analogous counting with an odd number of $v$'s returns 120 elements:
\begin{equation}\label{4pt:lbl}
    \underbrace{v^\mu \,v^\nu \,v^\rho \,n_i^\sigma}_{3, \atop 4\,\text{permutations}}\,, \qquad 
    \underbrace{v^\mu \,n_i^\nu \,n_j^\rho \,n_k^\sigma}_{3^3=27, \atop 4 \,\text{permutations}}\,.
\end{equation}
The simplest case is Figure~\ref{fig:4pt-fct}e with a vector and three pseudoscalars like the $\gamma\to 3\pi$ amplitude, or an axialvector with three scalars,
because here there is only one element: $v^\mu$.

\subsection{Correlation functions with $n>4$}

\begin{figure}[t]
\includegraphics[width=0.45\textwidth]{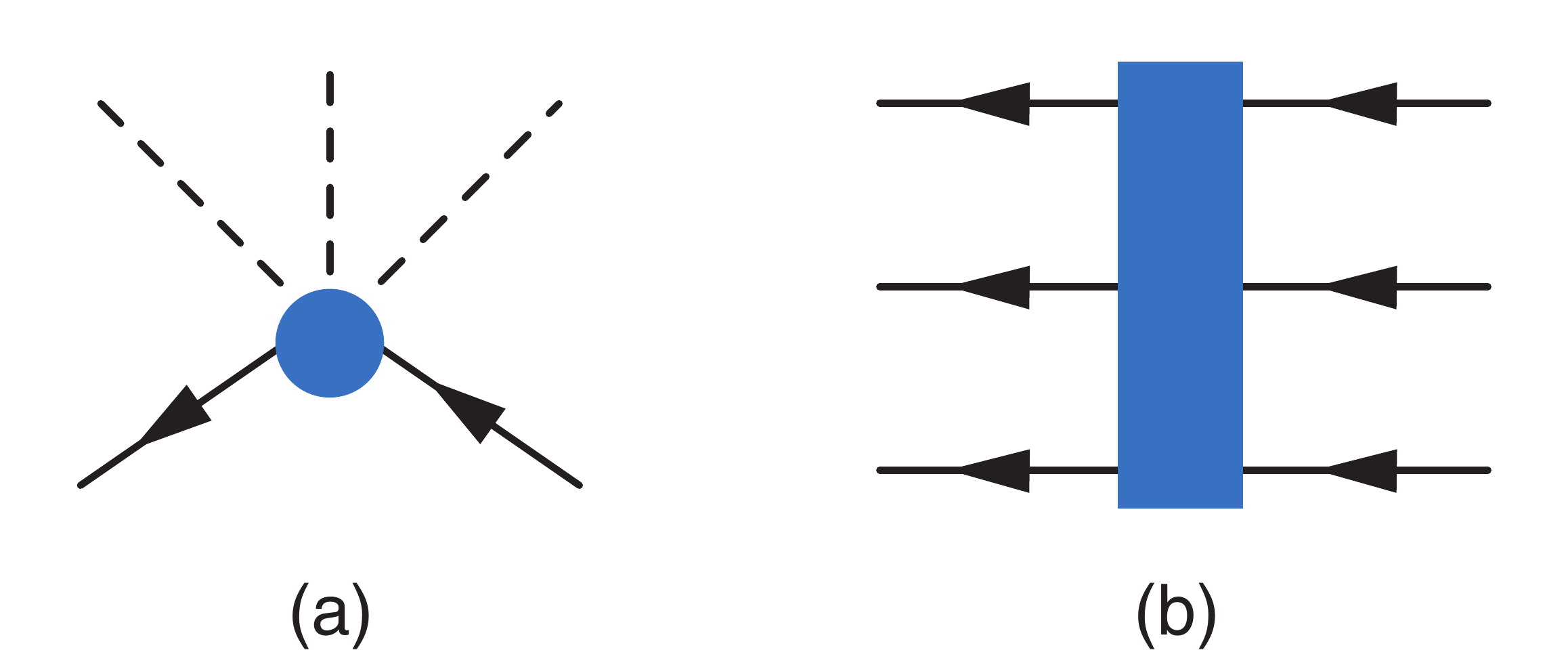}
\caption{Examples of five- and six-point functions. Solid lines represent fermions and dashed lines  scalars or pseudoscalars. \label{fig:56pt-fct}}
\end{figure}

Finally, what about $n$-point functions with $n > 4$? In that case we can orthogonalize all four vectors to arrive at
$n_1$, $n_2$, $n_3$, $n_4$. This entails
\begin{equation}\label{v-relations-2}
\begin{split}
   \delta^{\mu\nu} = \sum_{i=1}^4 n_i^\mu\,n_i^\nu\,,  \qquad 
   \gamma^\mu = \sum_{i=1}^4 n_i^\mu\,\slashed{n}_i\,, \qquad
   \slashed{n}_1\,\slashed{n}_2\,\slashed{n}_3\,\slashed{n}_4 = -\varepsilon \gamma_5\,, 
\end{split}
\end{equation}
where $\varepsilon$ is the pseudoscalar
\begin{equation}
    \varepsilon = \varepsilon^{\mu\nu\rho\sigma} n_1^\mu\,n_2^\nu\,n_3^\rho\,n_4^\sigma\,.
\end{equation}
In this case even $\gamma_5$ becomes redundant! For example, a five-point function with a fermion,  antifermion and three scalar legs
(Figure~\ref{fig:56pt-fct}a) has 16 tensors:
\begin{equation}
    \Omega' = \big\{ \underbrace{1}_{1}\,, \; \underbrace{\slashed{n}_i}_{4}\,, \; \underbrace{\slashed{n}_i\,\slashed{n}_j}_{6}\,, \; \underbrace{\slashed{n}_i\,\slashed{n}_j\,\slashed{n}_k}_{4}\,, \;
    \underbrace{\slashed{n}_i\,\slashed{n}_j\,\slashed{n}_k \,\slashed{n}_l}_{1} \big\}, \qquad i < j < k < l\,.
\end{equation}
A fermion six-point function like in Figure~\ref{fig:56pt-fct}b has 4096 tensors:
\begin{equation}
    \underbrace{\Omega'_{\alpha\beta}\,\Omega'_{\gamma\delta}\,\Omega'_{\lambda\tau}}_{16^3} \,.
\end{equation}
Swapping  parity now simply amounts to a multiplication with $\varepsilon$, otherwise it does not change the basis.

For $n>4$, the number of basis elements is therefore identical to what follows  from counting Dirac and Lorentz indices.
A tensor $\Gamma_{\alpha\beta\dots}^{\mu\nu\dots}$ with $2n_f$ Dirac indices (they must come in pairs) and $n_v$ Lorentz indices has $4^{2n_f+n_v}$ entries.
This is the upper limit, there cannot be more basis elements than there are entries in the matrix.
An $n>4$ point function with $n=2n_f+n_v+n_s$, where $n_f$ is the number of fermion pairs, $n_v$ the number of vector legs and $n_s$ the number of scalar legs,
also has $4^{2n_f+n_v}$ basis elements, so it meets that upper limit. For $n \leq 4$ the number is below the limit.  
Here one can also see that the number of basis elements grows exponentially with the number of fermion and vector legs.

\unskip

\subsection{Transversality}

$n$-point functions and matrix elements can be subject to further constraints. 
Amplitudes with gauge-boson legs satisfy gauge invariance through
Ward-Takahashi or Slavnov-Taylor identities.
As a consequence, they are either transverse with respect to each gauge-boson momentum in the corresponding Lorentz index,
or at least their non-transverse parts are constrained.
At the level of the basis construction with unit vectors, the contraction of each Lorentz index with a projector transverse
to its momentum is equivalent to removing one unit vector as a whole
from the Lorentz parts of the basis elements.
For instance in the fermion-vector vertex~\eqref{3pt:fv-vertex}, if $n_4^\mu$ is the normalized
four-momentum of the gauge boson, then the vector $n_4^\mu$ is longitudinal. Therefore, the transverse part consists of eight tensors only,
\begin{equation} 
     \left\{ \gamma_\perp^\mu, \; n_3^\mu \right\} \times \left\{  1\,, \; \slashed{n}_3\,, \; \slashed{n}_4\,, \; \slashed{n}_3\,\slashed{n}_4 \right\} .
\end{equation}
Here, $\gamma_\perp^\mu$ is the transverse projection of $\gamma^\mu$ with respect to $n_4$.
At least for counting purposes, the same strategy also works for other amplitudes.
The three-gluon vertex in Eq.~\eqref{3pt:3gv} has $2^3 + 3\cdot 2 = 14$ tensors in general and  $1^3 + 3\cdot 1 = 4$ transverse tensors.
The fermion Compton vertex in Eq.~\eqref{4pt:fcompton} has $(1 + 3^2 + 2\cdot 3)\cdot 8 = 128$ tensors in general and $(1 + 2^2 + 2\cdot 2)\cdot 8 = 72$ transverse tensors.
The light-by-light amplitude and the four-gluon vertex in Eq.~\eqref{4pt:lbl} have $1 + 6\cdot 3^2 + 3^4 = 136$ tensors in general and $1 + 6\cdot 2^2 + 2^4 = 41$ transverse tensors.

\subsection{Onshell constraints}\label{sec:onshell-1}

So far we considered the most general case where all external particles in an $n$-point function are offshell.
This is appropriate for amplitudes with quarks, gluons, virtual photons, etc., which are not  onshell physical states.
For onshell particles like electrons, nucleons or pions, the number of basis tensors is further reduced.
If  $n_i$ is the unit vector corresponding to the onshell fermion momentum and $u$ the respective Dirac spinor, 
then the Dirac equation $\slashed{n}_i\,u = u$ implies that all basis elements containing $\slashed{n}_i$ become redundant
upon contraction with Dirac spinors. In practice, it is often more convenient to attach a positive-energy projector $\Lambda = (1+\slashed{n}_i)/2$
to the fermion leg instead of a Dirac spinor, which has the same effect (since $\Lambda u = u$) but preserves the Dirac structure of the amplitude.
In any case, this entails that taking the fermion legs onshell is equivalent to removing $\slashed{n}_i$ from the basis for each onshell momentum.

For example, attaching Dirac spinors to the left and right of the fermion-scalar vertex in Eq.~\eqref{tau_i-sc}  amounts to 
removing all instances of $\slashed{n}_3$ and $\slashed{n}_4$, which leaves only one basis element (namely 1), like for 
the coupling of the nucleon to a scalar particle. The same happens for the coupling to a pseudoscalar particle, which leaves only $\gamma_5$.
For the onshell fermion-vector vertex, only two tensors remain ($\gamma^\mu$, $n_3^\mu$), which can be recast in 
the standard form in terms of a Dirac and Pauli form factor (see Eq.~\eqref{emff}  below).

For the onshell fermion-two-scalar vertex~\eqref{f2s-vertex}, the set $\Omega$ reduces to two tensors, $1$ and $\slashed{n}$, where $n$ is the remaining 
momentum in the four-point function that cannot be written as a linear combination of incoming and outgoing fermion momenta.
As a consequence, the onshell fermion four-point function~\eqref{f-4pt-fct} depends only on $2^2+2^2=8$ basis elements.
Because the onshell constraints also reduce the number of Lorentz invariants, symmetries like charge conjugation
and the Pauli principle can eliminate further tensors, namely those that are odd under the respective symmetry operation
and need to be multiplied with a Lorentz invariant that vanishes onshell (see also the discussion in Sec.~\ref{sec:onshell-2} below). 
This is the reason why the onshell nucleon-nucleon
scattering amplitude depends only on five tensors and not on eight.
For an onshell fermion Compton vertex~\eqref{4pt:fcompton} like the nucleon Compton scattering amplitude, 
one has $(1 + 2^2 + 2\cdot 2)\cdot 2 = 18$ transverse tensors.

\section{Tensor bases and symmetries}\label{sec:tensorbases}

While the tensor bases constructed in Sec.~\ref{sec:tensors} are very useful for counting purposes and also for solving dynamical equations,
they are usually not the best choice for discussing the properties of the   resulting dressing functions.
The reason is twofold. 
First, orthonormal bases  
do not play well with symmetries, and symmetry properties that would be obvious in bases where the symmetry is implemented
may look very obscure in an orthonormal basis.
Second, due to the orthonormalization the dressing functions usually have kinematic singularities,
kinematic zeros or at least kinematic dependencies.
In our context, \textit{kinematic singularities} refers to singularities in the dressing functions 
that appear only due to the choice of tensor basis but which
are not present in the $n$-point function as a whole.
These must be distinguished  from \textit{physical} or \textit{dynamical} singularities, which are actual poles and cuts induced by the dynamics
(like those discussed in Sec.~\ref{sec:2-2sc}).
Since the dressing functions contain all information about the $n$-point function, it is 
desirable to get rid of the kinematic singularities to reveal the actual dynamics in the process.

\begin{figure}[t]
\includegraphics[width=1\textwidth]{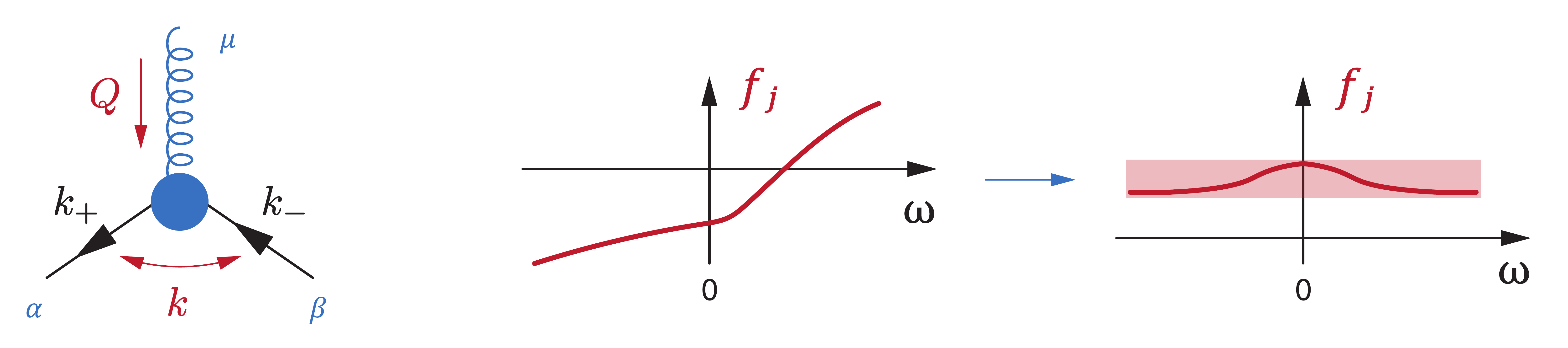}
\caption{Left: Kinematics in the fermion-vector vertex. Right: Implementing charge-conjugation symmetry
         simplifies the angular dependence of the dressing functions.\label{fig:fermion-vector}}
\end{figure}

Let us discuss these points again on the basis of a three-point function, namely the fermion-vector vertex shown in Figure~\ref{fig:fermion-vector}.
This object has many practical uses, e.g., in the form of the electron- or muon-photon vertex (which encode
the anomalous magnetic moments of the electron or muon), 
or the nucleon-photon vertex (which encodes the nucleon electromagnetic form factors),
or the quark-photon vertex or quark-gluon vertex. 
We label the momenta by $k_+$ (outgoing fermion momentum), $k_-$ (incoming fermion momentum),  and $Q = k_+ - k_-$ (incoming momentum of the vector particle).
We could work with $k_+$ and $k_-$ as the two independent momenta in the system or, since it is usually simpler,
the total momentum $Q$ and relative momentum $k$ defined by
\begin{equation}\label{kinematics-1}
   k = \frac{k_+ + k_-}{2}\,,  \quad Q = k_+ - k_-     \quad \Leftrightarrow \quad   k_\pm = k \pm \frac{Q}{2}  \,.
\end{equation} 

The vertex is Lorentz-covariant  and can be written as
\begin{equation}
   \Gamma_{\alpha\beta}^\mu(k,Q) = i \sum_{j=1}^{12} f_j(k^2, k\cdot Q, Q^2) \,(\tau_j)_{\alpha\beta}^\mu(k,Q)\,.
\end{equation}
The $f_j$ are the Lorentz-invariant dressing functions which depend on  
the three independent Lorentz invariants $k^2$, $k\cdot Q$ and $Q^2$.
The $\tau_j^\mu$ are the basis tensors, which carry all the Lorentz and Dirac structure. 
Therefore, once you know all twelve dressing functions in their full kinematic domain, you know everything there is to know about the vertex.

\subsection{Tensor basis}

From Eq.~\eqref{3pt:fv-vertex} we already know that the fermion-vector vertex has twelve tensors.
Let us  undo the orthogonalization and start from a plain tensor basis:
\begin{equation}\label{fv-basis-1}
\begin{array}{l@{\qquad}l@{\qquad}l@{\qquad}l}
   \gamma^\mu, & \gamma^\mu \,\slashed{k}\,, & \gamma^\mu \,\slashed{Q}\,, & \gamma^\mu \, \slashed{k} \,\slashed{Q}\,, \\
   k^\mu, & k^\mu \slashed{k}\,, & k^\mu \slashed{Q}\,, & k^\mu \slashed{k} \,\slashed{Q}\,, \\
   Q^\mu, & Q^\mu \slashed{k}\,, & Q^\mu \slashed{Q}\,, & Q^\mu \slashed{k} \,\slashed{Q}\,. \\
\end{array}
\end{equation}
This is a complete basis. But is it also the \textit{best} choice?
You could proceed by calculating the vertex with your preferred method and plot the resulting twelve dressing functions $f_i(k^2,k\cdot Q,Q^2)$
as functions of their three variables. However,  you would likely find the results unilluminating:
the $f_i$ would have complicated dependencies on all  variables,
and it would be difficult to identify which components actually drive the physics.
To analyze the system and gain insight in the underlying mechanisms, Eq.~\eqref{fv-basis-1} is far from ideal.
The good news is that symmetries are here to help!

\unskip

\subsection{Charge-conjugation symmetry}\label{sec:cc}

First of all, the fermion-vector vertex has a charge-conjugation symmetry:
\begin{equation}\label{cc-sym}
   \Gamma^\mu(k,Q) \stackrel{!}{=} -C \,\Gamma^\mu(-k,Q)^T C^T\,.
\end{equation}
Here, the superscript $T$ denotes a Dirac matrix transpose and $C = \gamma^4 \gamma^2$ is the charge-conjugation matrix, which satisfies
\begin{equation}
   C^T = C^\dag = C^{-1} = -C\,, \qquad C \,\gamma_5^T \, C^T = \gamma_5\,, \qquad C \,\gamma_\mu^T \,C^T = -\gamma_\mu\,.
\end{equation}
Eq.~\eqref{cc-sym} says that if we reverse the momenta ($k_+ \leftrightarrow -k_-$, which amounts to $k \leftrightarrow -k$)
and  spin lines (by applying the charge-conjugation matrix $C$), the vertex picks up a minus because it transforms like a vector.

Not all tensors in Eq.~\eqref{fv-basis-1} satisfy this constraint. Some of them do, like $\gamma^\mu$ and $k^\mu$,
\begin{equation}
    \gamma^\mu \; \to \; -C \,\gamma_\mu^T\,C^T = \gamma^\mu\,, \qquad
    k^\mu \; \to \; -C \,(-k^\mu)\, C^T = k^\mu\,,
\end{equation}
but others do not, like $Q^\mu$ which has the wrong $C$-parity:
\begin{equation}
    Q^\mu \; \to \; -C \,Q^\mu C^T = -Q^\mu\,.
\end{equation}
For some other tensors it is not just a matter of plus or minus, like
\begin{equation}
   \gamma_\mu\,\slashed{Q} \; \to \; -C\,\slashed{Q}^T \gamma_\mu^T\,C^T = \slashed{Q}\,C\,\gamma_\mu^T\,C^T = -\slashed{Q}\,\gamma_\mu = \gamma_\mu\,\slashed{Q} - 2Q^\mu\,,
\end{equation}
which  would lead to cumbersome symmetry relations between the dressing functions.

Obviously, things could become simpler if we implemented the symmetry directly in the basis. 
For example, if we attach a factor $k\cdot Q$ to $Q^\mu$ we have the correct $C$ parity: 
\begin{equation}
   k\cdot Q\,Q^\mu \; \to \; + k\cdot Q\,C\,Q^\mu C^T  = k\cdot Q\,Q^\mu\,.
\end{equation}
For $\gamma^\mu\,\slashed{Q}$ we can instead use a commutator:
\begin{equation}
   [\gamma_\mu, \slashed{Q}] \; \to \; -C\,[ \slashed{Q}^T, \gamma_\mu^T ] \,C^T = -[ C\,\slashed{Q}^T C^T, C\,\gamma_\mu^T\,C^T] = -[\slashed{Q}, \gamma_\mu] = [\gamma_\mu, \slashed{Q}]\,.
\end{equation} 
In short, we can enforce the charge-conjugation symmetry~\eqref{cc-sym} for each basis element individually if
we use commutators and attach factors $k\cdot Q$ where necessary.
Abbreviating $\omega = k\cdot Q$, this leads to the new basis
\begin{equation}\label{fv-basis-2}
\begin{array}{l@{\qquad}l@{\qquad}l@{\qquad}l}
   \gamma^\mu, & i\omega\,[\gamma^\mu, \slashed{k}]\,, & i[\gamma^\mu, \slashed{Q}]\,, & [\gamma^\mu, \slashed{k}, \slashed{Q}]\,, \\
   ik^\mu, & k^\mu \slashed{k}\,, & \omega\,k^\mu \slashed{Q}\,, & ik^\mu [\slashed{k} \,\slashed{Q}]\,, \\
   i\omega\,Q^\mu, & \omega\,Q^\mu \slashed{k}\,, & Q^\mu \slashed{Q}\,, & i\omega\,Q^\mu [\slashed{k} \,\slashed{Q}]\,. \\
\end{array}
\end{equation}
In the first line we used the triple commutator $[A,B,C] = [A,B] C + [B,C]A + [C,A]B$ as the totally antisymmetric combination of $A$, $B$ and $C$.
For convenience we also included $i$ factors to make all dressing functions real.

In this new basis, all tensors satisfy the constraint~\eqref{cc-sym} individually. 
But this means that also each dressing function must satisfy the constraint. Because $\omega = k\cdot Q$ is odd under charge conjugation,
this implies that the dressing functions can only depend on $\omega^2$. Thus the vertex reads
\begin{equation}\label{fv-vertex-1}
   \Gamma_{\alpha\beta}^\mu(k,Q) = i \sum_{i=1}^{12} f_i(k^2, \omega^2, Q^2) \,(\tau_i)_{\alpha\beta}^\mu(k,Q)\,,
\end{equation}   
where the $\tau_i^\mu$ are the tensors in Eq.~\eqref{fv-basis-2}.
This also simplifies the momentum dependence of the dressing functions, 
because now they must be even functions in $\omega$.
The situation is sketched in Figure~\ref{fig:fermion-vector}: In the original, `plain' basis
the dressing functions did not need to have a symmetry  in $\omega$, while in the new one
they must be even functions. The variable $\omega = k\cdot Q$ is the cosine of a hyperspherical angle,
so the dependence on $\omega$ encodes the angular dependence of the dressing functions.
An even function has less wiggle room than one that is neither even nor odd, and indeed it turns out that 
the angular dependencies are usually rather small or even flat. Thus, 
by choosing a basis that implements the symmetry we effectively eliminated a variable from the system!
Mathematically speaking, we have arranged the tensors and dressing functions
into singlets of the permutation group $S_2$.

\subsection{Gauge invariance}

There is  more we can say about the structure of the fermion-vector vertex.
Usually we are dealing with gauge theories, where the vector particle is a gauge boson like a photon, gluon, etc.
Amplitudes with gauge-boson legs are subject to gauge invariance, which is manifest through Ward-Takahashi identities (WTIs)
or Slavnov-Taylor identities (STIs). 
To keep things simple, let us start with a scalar-photon vertex.
In that case there are only two tensors, $k^\mu$ and $Q^\mu$. The $C$-parity relation~\eqref{cc-sym}
reduces to $\Gamma^\mu(k,Q) = -\Gamma^\mu(-k,Q)$,  
and the analogue of Eqs.~(\ref{fv-basis-2}--\ref{fv-vertex-1}) is 
\begin{equation}\label{sc-vertex-1}
   \Gamma^\mu(k,Q) = c_1\,k^\mu + c_2\,\omega Q^\mu\,,
\end{equation}
where the dressing functions $c_i$ depend on $k^2$, $\omega^2$ and $Q^2$.
In the following we will assume that the $c_i$ are regular at $Q\to 0$, so that the limit $\Gamma^\mu(k,0)$ is well-defined.
The vertex is subject to the WTI
\begin{equation}
     Q^\mu \Gamma^\mu(k,Q) = D(k_+)^{-1} - D(k_-)^{-1}\,,
\end{equation}
where $D(k)$ is  the propagator of the scalar particle and $D(k)^{-1}$ its inverse.
In the limit $Q^\mu \to 0$, this reduces to the Ward identity
\begin{equation}
     \Gamma^\mu(k,0) = \frac{d D(k)^{-1}}{dk^\mu}\,. 
\end{equation}
One can see that the WTI fixes the longitudinal part (with respect to the total momentum $Q$) of the vertex  but leaves its transverse part 
unconstrained. How does this affect the dressing functions in Eq.~\eqref{sc-vertex-1}?

To condense the notation a bit, we define the difference quotient
\begin{equation}
    \Delta(k,Q) := \frac{D(k_+)^{-1} - D(k_-)^{-1}}{k_+^2 - k_-^2}\,, \qquad
    \Delta(k,0) = \frac{dD(k)^{-1}}{dk^2}\,.
\end{equation}
For a free propagator of the form $D(k) = 1/(k^2 + m^2)$ we would simply have $\Delta(k,Q) = 1$,
but let us keep things general.

From Eq.~\eqref{kinematics-1} we have $k_+^2 - k_-^2 = 2k\cdot Q = 2\omega$, so we can write the WTI and Ward identity as
\begin{equation}\label{wti-sc-2}
    Q^\mu \Gamma^\mu(k,Q) = 2\omega\,\Delta(k,Q)\,, \qquad \Gamma^\mu(k,0) = 2k^\mu \Delta(k,0)\,.
\end{equation}
Plugged back into Eq.~\eqref{sc-vertex-1}, this yields $c_1 = 2\Delta - c_2 \,Q^2$ and hence the result
\begin{equation}\label{sc-vertex-gt}
   \Gamma^\mu(k,Q) = 2\Delta\,k^\mu - c_2\left( Q^2 \,k^\mu - \omega\,Q^\mu\right) = 2\Delta\,k^\mu - c_2\,t^{\mu\nu}_{QQ} \,k^\nu\,, 
\end{equation}
where for later purposes we defined
\begin{equation}\label{tAB}
   t^{\mu\nu}_{AB} = A\cdot B\,\delta^{\mu\nu} - B^\mu A^\nu\,.
\end{equation}

Because this example is so simple, it is also a good testing ground for discussing how \textit{not} to do it. 
Let us first define the usual transverse and longitudinal projectors with respect to the photon momentum $Q$ by
\begin{equation}
    T^{\mu\nu}_Q = \delta^{\mu\nu} - \frac{Q^\mu Q^\nu}{Q^2}\,, \qquad
    L^{\mu\nu}_Q = \frac{Q^\mu Q^\nu}{Q^2}\,.
\end{equation}
Since they sum up to $\delta^{\mu\nu}$, we can equally write the vertex as
\begin{equation}\label{fv-vertex-split-tl}
   \Gamma^\mu(k,Q) = T^{\mu\nu}_Q \Gamma^\nu(k,Q) +  L^{\mu\nu}_Q \Gamma^\nu(k,Q) = \Gamma^\mu_\perp(k,Q) + \frac{2\Delta(k,Q)}{Q^2} \, \omega\, Q^\mu.
\end{equation}
From Eq.~\eqref{sc-vertex-1} the transverse projection of the vertex is $\Gamma^\mu(k,Q) = c_1\,k^\mu_\perp$,
where $k^\mu_\perp = T^{\mu\nu}_Q k^\nu$ is the transverse projection of the momentum $k$ with respect to the total momentum $Q$.
This  decomposition  obviously satisfies the WTI~\eqref{wti-sc-2}, and it is a valid tensor decomposition with a  transverse tensor $k^\mu_\perp$
and a  longitudinal tensor $\omega Q^\mu$.
However, the longitudinal dressing function has a kinematic singularity at $Q^2 = 0$, which
was not present in the original vertex and only came about by our  choice of tensor basis.
As such it must cancel with the transverse part, which it does because $k^\mu_\perp$ also has a $1/Q^2$ singularity.
Thus, the transverse and longitudinal parts are kinematically related.
This can also be seen by the fact that
the longitudinal part alone violates the Ward identity even though it satisfies the WTI;
the Ward identity relates the transverse and longitudinal parts.
In general, a separation into \textit{transverse} and \textit{longitudinal} pieces  induces kinematic singularities. 

We should emphasize  that there is nothing \textit{wrong} with kinematic singularities, they just obscure the physical interpretation.
If the $1/Q^2$ singularity in Eq.~\eqref{fv-vertex-split-tl} were dynamical, it could imply a massless particle.
Such singularities may very well occur~\cite{Aguilar:2011xe,Aguilar:2016ock,Eichmann:2021zuv,Aguilar:2021uwa,Aguilar:2022thg}. However, the one in Eq.~\eqref{fv-vertex-split-tl} is just a kinematic singularity.
How can we get rid of it?  
Eq.~\eqref{sc-vertex-gt} shows that there is a way to separate the transverse part from the rest without introducing kinematic singularities.
Suppose we did not know anything about the WTI. In that case, we could work out the transverse part alone by solving the condition
\begin{equation}
   Q^\mu \Gamma^\mu(k,Q) = \omega\,(c_1 + c_2\,Q^2) \stackrel{!}{=} 0\,.
\end{equation}
To avoid kinematic singularities, we need to solve this condition for $c_1$ and not $c_2$, which entails $c_1 = -c_2\,Q^2$.
Plugged into Eq.~\eqref{sc-vertex-1}, this yields the transverse part $-c_2\,t^{\mu\nu}_{QQ}\,k^\nu$, which does not have any $1/Q^2$ terms.
(Had we solved for $c_2$ instead, we would  get $c_1\,k^\mu_\perp$ again.)
To obtain the full vertex, since we eliminated the coefficient $c_1$ we must put back its corresponding tensor $k^\mu$ to arrive at
\begin{equation}
   \Gamma^\mu(k,Q) = g\,k^\mu -c_2\,t^{\mu\nu}_{QQ} \,k^\nu \,.
\end{equation}
This is  a kinematically independent separation into a `gauge part' and a transverse part, 
which means that the dressing functions $g$ and $c_2$ do not know anything about each other
and one could remove one or the other term without running into problems.

The gauge part $g\,k^\mu$ is not longitudinal but still constrained by gauge invariance.
If we \textit{now} work out the WTI, we find $g = 2\Delta$ and arrive at the final result in Eq.~\eqref{sc-vertex-gt}.
The gauge part alone satisfies both the WTI and the Ward identity,
and it is completely determined by the scalar propagator.
At tree level, the propagator is simply $D(k) = 1/(k^2 + m^2)$ and therefore $\Delta = 1$,
so the tree-level vertex is $2k^\mu$. Thus, the structure information rests partially in 
$\Delta$, which reflects the change in the propagator,
and in the dressing function $c_2(k^2,\omega^2,Q^2)$, which is not constrained by gauge invariance and reflects the genuine dynamics.
Such a separation into a gauge part and a transverse part which is free of any kinematic constraints 
is also called a \textit{minimal basis}.

\begin{figure}[t]
\includegraphics[width=1\textwidth]{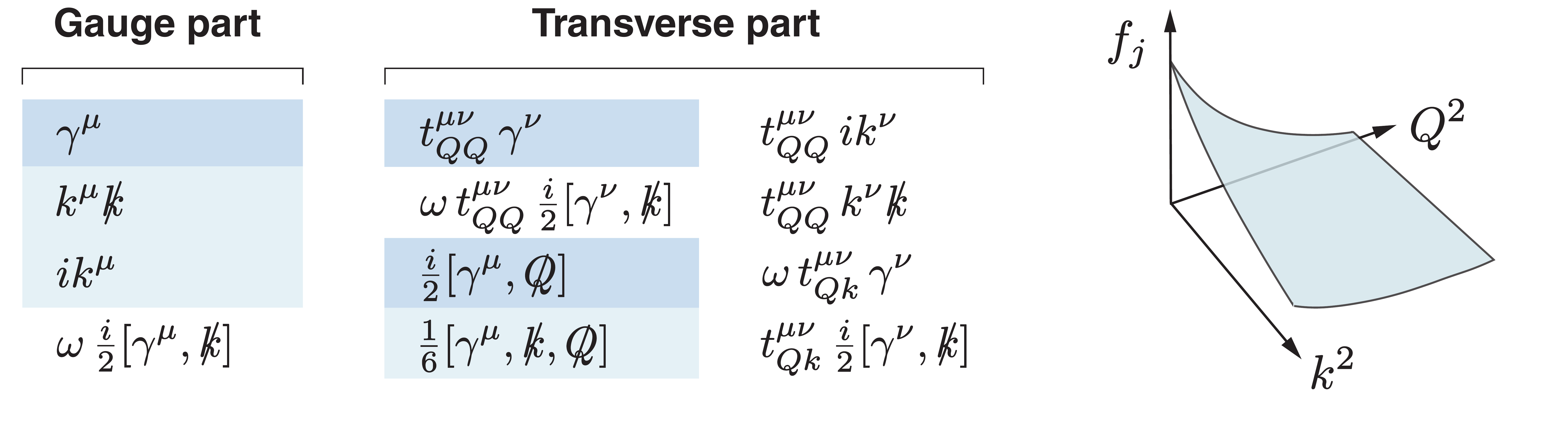}
\caption{Left: Tensor basis for the fermion-vector vertex in Eq.~\eqref{fv-vertex-final} with gauge part ($G_1 \dots G_4$)
         and transverse part ($T_1 \dots T_4$ in the left column, $T_5 \dots T_8$ in the right column). 
         The tensors with colored background are those with the lowest momentum powers.
         Right: 
         Once the \textit{kinematic fog} is removed, the dressing functions become simple and their momentum
         dependence is governed by dynamical singularities in the timelike region.\label{fig:fermion-vector-2}}
\end{figure}   
\unskip

The same strategy can be applied to the fermion-photon vertex~\cite{Ball:1980ax,Kizilersu:1995iz,Davydychev:2000rt,Skullerud:2002ge,Bashir:2011dp,Eichmann:2012mp}.
Starting from the vertex (\ref{fv-basis-2}--\ref{fv-vertex-1}), 
one can first work out the transversality condition $Q^\mu \Gamma^\mu(k,Q) = 0$.
This yields four equations, which must be solved   without introducing kinematic singularities. 
Plugged back into the vertex, one arrives at eight transverse tensors.
The complementary tensors from the gauge part belong to those four coefficients that were eliminated.
The result is
\begin{equation}\label{fv-vertex-final}
   \Gamma^\mu(k,Q) = i \left[ \sum_{j=1}^4 g_j\,G_j^\mu + \sum_{j=1}^8 f_j\,T_j^\mu \right],
\end{equation}
where the $G_j^\mu$ and $T_j^\mu$ are the tensors from the gauge and transverse part, respectively.
They are collected in Figure~\ref{fig:fermion-vector-2}. The quantity $t^{\mu\nu}_{AB}$ is defined in Eq.~\eqref{tAB}
and the triple commutator below Eq.~\eqref{fv-basis-2}. 
Because $t^{\mu\nu}_{AB}$ is transverse to
the momentum $A$ in the index $\mu$ and transverse to $B$ in $\nu$, the transversality with respect to 
the total momentum $Q$ in the index $\mu$ is manifest in the basis. 
Once again, there are no kinematic singularities and all dressing functions
$g_j$ and $f_j$ are kinematically independent. All tensors share the same $C$ parity as the vertex 
so that the dressing functions depend on $k^2$, $\omega^2$ and $Q^2$.

Finally, we can also work out the WTI for the gauge part to fix the four $g_i$.
Taking for example the fermion-photon vertex, it is subject to the vector WTI
\begin{equation}
   Q^\mu \Gamma^\mu(k,Q) = S(k_+)^{-1} - S(k_-)^{-1}\,,
\end{equation}
where 
\begin{equation}
   S(k)^{-1} = A(k^2)\left(i\slashed{k} + M(k^2)\right) = A(k^2)\,i\slashed{k} + B(k^2)
\end{equation}
is the inverse fermion propagator including the fermion mass function $M(k^2)$.
Applying the WTI yields the Ball-Chiu vertex~\cite{Ball:1980ax}
\begin{equation}
   g_1 = \Sigma_A\,, \quad g_2 = 2\Delta_A\,, \quad g_3 = -2\Delta_B\,, \quad g_4 = 0\,,
\end{equation}
where
\begin{equation}
   \Sigma_A = \frac{A(k_+^2) + A(k_-^2)}{2}\,, \quad  
   \Delta_A = \frac{A(k_+^2) - A(k_-^2)}{k_+^2 - k_-^2}\,, \quad  
   \Delta_B = \frac{B(k_+^2) - B(k_-^2)}{k_+^2 - k_-^2} 
\end{equation}
are the averages and differences of the quark propagator dressing functions.
For a tree-level propagator one has $A(k^2) = 1$ and $M(k^2) = m$ and thus $\Sigma_A = 1$, $\Delta_A = \Delta_B = 0$,
so the tree-level vertex is simply $i\gamma^\mu$. Everything else is  dynamics: on the one hand, 
the dynamics from the propagator which transforms $i\gamma^\mu$ into the Ball-Chiu vertex,  
and on the other hand the  dynamics from the vertex through its transverse part.

Since  now we have cleared our view from the  \textit{kinematic fog}, 
the momentum dependence of the dressing functions $f_j$
is governed by actual, dynamical singularities.
Because a photon can fluctuate into vector mesons, the transverse part of the quark-photon vertex 
must have vector-meson poles at timelike momenta $Q^2 = -m_{V_i}^2$, 
which  appear in every dressing function $f_j$ and are reproduced in dynamical calculations~\cite{Maris:1999bh,Maris:2005tt,Krassnigg:2004if,Eichmann:2016yit,Eichmann:2019bqf,Miramontes:2021xgn}.
Since they further propagate into the electromagnetic form factors of hadrons, this is the underlying origin of vector-meson dominance.

Another useful feature of the basis in Figure~\ref{fig:fermion-vector-2} is that it 
allows for a power counting. Each momentum $k$ or $Q$ contributes one momentum power and $\omega = k\cdot Q$ two.
The tensors with the lowest momentum powers (zero, one or two) are highlighted in color in Figure~\ref{fig:fermion-vector-2}, 
while the remaining ones have three, four or five momentum powers.
This is analogous to the derivative expansions in effective field theories, since higher derivatives in the Lagrangian
translate to tensors with higher momentum powers.
Those tensors are suppressed at low momenta, so their contributions become less important in the infrared.
Moreover, their dressing functions usually fall off faster in $k^2$ or $Q^2$, so their
contributions also become less important in the ultraviolet. 
Therefore, the most important tensors are usually those with the lowest momentum powers. 

In fact, the basis in Figure~\ref{fig:fermion-vector-2} is the one with the lowest possible overall momentum powers:
the power zero appears only once ($\gamma^\mu$), one appears twice ($ik^\mu$, $\frac{i}{2}[\gamma^\mu,\slashed{Q}]$), two and three appear three times each,
four  twice and five once. Written out explicitly, if $\#$ denotes the instances of tensors with momentum powers $0, 1, 2, 3, 4, \dots$, then
\begin{equation}
   \# = \{1,2,3,3,2,1,0,0,0,\dots\}\,.
\end{equation}
These are the same powers appearing in the original basis~\eqref{fv-basis-2},
which means the rearrangement based on gauge symmetry was possible without any kinematic divisions.
Therefore, it is also not possible to find a basis with lower momentum powers that is still free of kinematic constraints.
We can always replace tensors with new ones by attaching powers of $Q^2$, $k^2$ or $\omega^2$, but this will
increase the momentum powers.
In this sense,  a minimal basis is one with the lowest possible momentum powers that makes the gauge symmetry explicit.
Note that for this statement to make sense, it was necessary to  arrange the tensors into singlets under charge conjugation 
because the prefactors $\omega =k\cdot Q$ also contribute to the counting, i.e., the permutation-group symmetries
must be worked out beforehand.
For the scalar-vector vertex in Eqs~\eqref{sc-vertex-1} and~\eqref{sc-vertex-gt}  the analogous counting is
\begin{equation}
   \# = \{0,1,0,1,0,0,0,\dots\}\,.
\end{equation}

There are three tensors in the fermion-photon vertex that do not depend on the relative momentum $k^\mu$ at all: $\gamma^\mu$, $t^{\mu\nu}_{QQ}\,\gamma^\nu$,
and $\frac{i}{2}[\gamma^\mu,\slashed{Q}]$ related to $g_1$, $f_1$ and $f_3$, respectively.   
If one of their dressing functions were much smaller than expected,
this would point towards interesting dynamics. There are indeed indications from  functional and lattice calculations of the quark-photon vertex that
the dressing function $f_3$ associated with the `anomalous magnetic moment'  of the quark is very small~\cite{Eichmann:2014qva,Leutnant:2018dry}.

\subsection{Onshell vertices}\label{sec:onshell-2}

Let us briefly return to Sec.~\ref{sec:onshell-1} and the discussion of the onshell vertices.
When we consider the scalar-photon vertex  either from Eq.~\eqref{sc-vertex-1} or~\eqref{sc-vertex-gt},
if the scalar particle is onshell it satisfies $k_\pm^2 = -m^2$ and hence
$k^2 = -m^2 - Q^2/4$ and $\omega=0$. Therefore, only $Q^2$ survives as an independent variable,
and all $\omega$ factors in the basis are zero. In Eq.~\eqref{sc-vertex-gt} 
this amounts to the replacement $t^{\mu\nu}_{QQ}\,k^\nu \to Q^2\,k^\mu$.
Moreover, the  propagator of an onshell particle is $D(k^2) = 1/(k^2+m^2)$, which implies $\Delta = 1$ and therefore
\begin{equation}
   \Gamma^\mu(k,Q) \Big|_\text{onshell} = 2k^\mu\left( 1 - \frac{c_2}{2}\,Q^2\right) =: 2k^\mu F(Q^2)\,.
\end{equation}
This is the general form of a scalar or pseudoscalar electromagnetic current matrix element, with an electromagnetic form factor $F(Q^2)$ 
that is constrained by $F(0) = 1$ by means of the electromagnetic WTI.

The same applies to a fermion-photon vertex in either of the decompositions~\eqref{fv-basis-2} or~\eqref{fv-vertex-final};
all basis elements proportional to $\omega$ are zero. 
Furthermore, the  contraction
with onshell spinors $\conjg{u}(k_+)$ and $u(k_-)$ (or positive-energy projectors) on the left and right 
makes all instances of $\slashed{k}$ and $\slashed{Q}$ redundant, because their linear combinations $\slashed{k}_+$ and $\slashed{k}_-$
become redundant by the Dirac equations 
\begin{equation}
   \slashed{k}_- u(k_-) = im \,u(k_-)\,, \qquad \conjg{u}(k_+) \slashed{k}_+ = im \,\conjg{u}(k_+)\,. 
\end{equation}
As a result, only $\gamma^\mu$ and $k^\mu$ survive.
Because the combination
\begin{equation}
   \gamma^\mu + \frac{ik^\mu}{m} - \frac{i}{4m}[\gamma^\mu,\slashed{Q}]\,,
\end{equation}
vanishes inside  the spinor contraction (this is the Gordon identity),
the onshell vertex can be written in the usual form
\begin{equation}\label{emff}
    \conjg{u}(k_+)\,\Gamma^\mu(k,Q)\,u(k_-) \Big|_\text{onshell} = i \conjg{u}(k_+)\left( F_1(Q^2) \gamma^\mu + F_2(Q^2)\,\frac{i}{4m}[\gamma^\mu, \slashed{Q}]\right) u(k_-) \,.
\end{equation}
The Dirac form factor $F_1(Q^2)$ at $Q^2=0$ returns the fermion's charge, while the Pauli form factor $F_2(Q^2)$ returns
its anomalous magnetic moment $F_2(0)$, which is not constrained by
gauge invariance.
Their relations with the $f_j$ can be found in Ref.~\cite{Eichmann:2018ytt}.

\subsection{Other examples}

The decomposition into gauge and transverse parts outlined above  
can also be applied to other $n$-point functions with gauge-boson legs.
Some cases are well-known, like the fermion-photon vertex discussed above~\cite{Ball:1980ax,Kizilersu:1995iz,Davydychev:2000rt,Skullerud:2002ge,Bashir:2011dp,Eichmann:2012mp} or the nucleon Compton scattering amplitude~\cite{Bardeen:1968ebo,Tarrach:1975tu,Drechsel:2002ar,Eichmann:2018ytt}.
The transverse part of the light-by-light scattering amplitude has been established in~\cite{Colangelo:2015ama,Eichmann:2015nra}.
Applications to transition form factors between  $J=\tfrac{1}{2}$ and $J=\tfrac{3}{2}$ baryons can be found in~\cite{Eichmann:2018ytt}.
There exist also cases where it is not possible to find a minimal basis, i.e., where one cannot avoid kinematic singularities 
when solving the transversality equations~\cite{Eichmann:2018ytt}.
If the procedure is possible, however, it can simplify 
the discussion of the dynamics significantly.

\newpage

\section{Which Lorentz invariants are \textit{best}?}\label{sec:li}

Returning again to the general structure of an $n$-point function in Eq.~\eqref{nptfct-gen},
we have now extensively discussed the aspect of tensor bases. What about the Lorentz invariants?
First of all, how many are there?

This can be directly read off from Eq.~\eqref{specific-frame}: The number of Lorentz invariants equals the number of independent entries in the vectors.
A two-point function ($n=2$) depends only on one vector, e.g. $p$, so there is only one Lorentz-invariant $p^2$.
A three-point function ($n=3$) depends on two vectors, e.g. $p$ and $q$, which gives three Lorentz invariants: $p^2$, $q^2$ and $p\cdot q$.
A four-point function ($n=4$) depends on three vectors, which gives six Lorentz invariants.
But then from $n=5$ onwards, the number only grows by 4 since we are limited by four spacetime dimensions.
For general $n\geq 4$ the number of Lorentz invariants is then $4n-10$, so it grows linearly with $n$.
If there are further onshell constraints, they will reduce that number by 1 for each constraint.

Actually this has some strange consequences. With $n-1$ vectors in the system, in principle there are
$1 + 2 + 3 + \dots + n-1 = n(n-1)/2$ possible Lorentz-invariant combinations. 
For example, for $n=6$ there are six independent vectors $p_1, \dots p_5$,
from where one can  write down $n(n-1)/2 = 15$ Lorentz invariants:
\begin{equation}
   \begin{array}{r @\qquad r @\qquad r @\qquad r @\qquad r}
      p_1^2\,,  & p_2^2\,,  & p_3^2\,,  & p_4^2\,,  & p_5^2\,,  \\
                & p_1\cdot p_2\,,  & p_1\cdot p_3\,, & p_1\cdot p_4\,, & p_1\cdot p_5\,,  \\ 
                &                  & p_2\cdot p_3\,, & p_2\cdot p_4\,, & p_2\cdot p_5\,,  \\  
                &                  &                 & p_3\cdot p_4\,, & p_3\cdot p_5\,,  \\  
                &                  &                 &                 & p_4\cdot p_5\,.
  \end{array}
\end{equation}
However, $4n-10 = 14$, so there must be a linear relation between them.  Therefore, for $n \geq 6$ even the Lorentz invariants become redundant!

Which Lorentz invariants are \textit{best}, i.e., most convenient to use?
Once again, symmetries are helpful. Consider a three-point function with
three incoming momenta $p_1$, $p_2$, $p_3$ and $p_1+p_2+p_3=0$.
If the three-point function has some symmetry, picking  $p_1$ and $p_2$ as independent momenta is probably not the best choice 
because the symmetry will lead to cumbersome relations between the Lorentz invariants $p_1^2$, $p_2^2$ and $p_1\cdot p_2$.
For example, a fermion-vector vertex like in Section~\ref{sec:cc} has a charge-conjugation symmetry which amounts to $p_1 \leftrightarrow p_2$
for the incoming momenta $p_1 = -k_+$ and $p_2 = k_-$,
i.e., it is subject to the permutation group $S_2$. This suggests using relative and total momenta like in Eq.~\eqref{kinematics-1}:
\begin{equation}
   k = \frac{p_2 - p_1}{2}\,, \qquad Q = p_3 = -(p_1+p_2)\,.
\end{equation}
The Lorentz invariants $k^2 = (p_1^2 + p_2^2 - 2p_1\cdot p_2)/4$ and $Q^2 = p_1^2 + p_2^2 + 2p_1\cdot p_2$ are then invariant under $p_1 \leftrightarrow p_2$ (i.e., they are $S_2$ singlets),
whereas $\omega = k\cdot Q = (p_1^2 - p_2^2)/2$ is antisymmetric (it is an antisinglet).
In particular, once all tensors are arranged into singlets, like we did in Eq.~\eqref{fv-basis-2} or Fig.~\ref{fig:fermion-vector-2},
their dressing functions must be singlets as well, so they can only depend on $k^2$, $Q^2$ and $\omega^2$.

\begin{figure}[t]
\begin{adjustwidth}{-\extralength}{0cm}
\includegraphics[width=18.0cm]{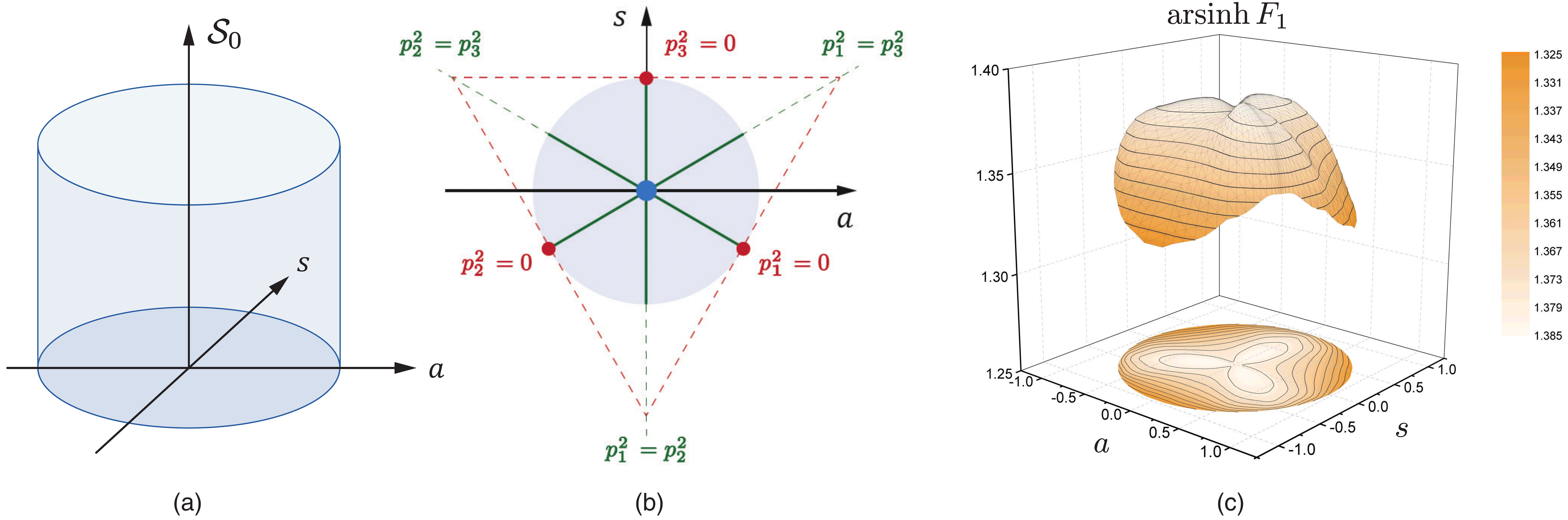}
\caption{(a) The spacelike kinematic domain relevant for the three-gluon vertex is a cylinder spanned by the variables $\mathcal{S}_0$, $a$ and $s$. 
(b) A slice at fixed $\mathcal{S}_0$ defines a Mandelstam plane in the variables $a$ and $s$; the lines at fixed $p_1^2$, $p_2^2$ and $p_3^2$
form a triangle. (c) Leading dressing function of the three-gluon vertex obtained from its Dyson-Schwinger equation at fixed $\mathcal{S}_0 = 102$~GeV$^2$~\cite{Eichmann:2014xya}. \label{fig:3gv}}
\end{adjustwidth}
\end{figure}

If the three-point function is fully symmetric like the three-gluon vertex, it has an $S_3$ symmetry.
In that case, it is useful to cast the Lorentz invariants into $S_3$ multiplets~\cite{Eichmann:2014xya}:
\begin{equation}
   \mathcal{S}_0 =  \frac{p_1^2 + p_2^2 + p_3^2}{6} = \frac{k^2}{3} + \frac{Q^2}{4}\,,  \qquad 
\begin{array}{rl}
   a &= \displaystyle\sqrt{3}\,\frac{p_2^2 - p_1^2}{p_1^2 + p_2^2 + p_3^2} = -\frac{k\cdot Q}{\sqrt{3}\,\mathcal{S}_0}\,, \\[6mm]
   s &= \displaystyle\frac{p_1^2 + p_2^2 - 2p_3^2}{p_1^2 + p_2^2 + p_3^2} = \frac{1}{\mathcal{S}_0}\left( \frac{k^2}{3} - \frac{Q^2}{4}\right).
\end{array}
\end{equation}
The symmetric variable $\mathcal{S}_0$ is a singlet under the permutation group $S_3$.
The remaining variables are mixed-antisymmetric ($a$) or mixed-symmetric ($s$) in the indices 1 and 2 and transform like a doublet under $S_3$.
The spacelike region defined by $k^2 > 0$, $Q^2 > 0$ and $(k\cdot Q)^2 < k^2 \,Q^2$ translates to $\mathcal{S}_0 > 0$ and the unit disk $a^2 + s^2 < 1$. 
Thus, the kinematic region is a cylinder with radius 1 as shown in Figure~\ref{fig:3gv}a. 

\unskip

For a fixed value of $\mathcal{S}_0$ one can draw a Mandelstam plane in the variables $a$ and $s$, which encodes the angular dependence (Figure~\ref{fig:3gv}b).
Like in the $2\to 2$ scattering example in Section~\ref{sec:2-2sc},
the Mandelstam plane shows the singularities in the process by straight lines.
In the three-gluon vertex these cannot be physical singularities because a gluon is not observable.
However, since an analytic function without any singularities would be a constant, there must be singularities \textit{somewhere}.
Due to the permutation symmetry, any of these singularities must line up symmetrically;
a singularity on one side of the Mandelstam plane must also appear on the other two sides.

If we employ a symmetric tensor basis for the three-gluon vertex such that every basis element  shares the Bose symmetry of
the full vertex, then the dressing functions must be fully symmetric.
They are then again invariant both under a 120$^\circ$ rotation and a left-right reflection of the Mandelstam plane.
This puts rather tight constraints on the kinematics and already tells us that the angular dependencies should be rather mild,
which  is indeed the case: Figure~\ref{fig:3gv}c shows the leading dressing function of the three-gluon vertex
at a scale $\mathcal{S}_0 = 102$ GeV$^2$. At this large momentum, the angular dependence is already sizeable
but still rather tame;   with lower $\mathcal{S}_0$ it becomes even smaller and the functions become more and more flat~\cite{Eichmann:2014xya}.
This behavior has been given the name \textit{planar degeneracy}~\cite{Aguilar:2023qqd}.
In practice it means that   the dressing functions $f_j(\mathcal{S}_0, a, s) \approx f_j(\mathcal{S}_0)$ essentially depend on the symmetric variable $\mathcal{S}_0$ only. Therefore, by working with a symmetric tensor basis, we   effectively eliminated two variables!

From this discussion it is clear that symmetries are very useful in dealing with higher $n$-point functions.
Once all symmetries have been implemented in the basis elements,
the dressing functions must be singlets. As a  consequence,  their angular dependencies are usually mild
and for low momenta even negligible. This can be used to construct high-quality approximations, which drastically
reduce the numerical demand while having little effect on the dynamics. 
Similar applications to four- and five-point functions
using the permutation groups $S_4$ and $S_5$ can be found in~\cite{Eichmann:2015nra,Eichmann:2025etg} 
and  have been  employed in different contexts such as, e.g., hadronic light-by-light scattering~\cite{Eichmann:2015nra,Colangelo:2017fiz},
the  process $\gamma^\ast \pi\to\pi\pi$~\cite{Miramontes:2025ofw}, 
the four-gluon vertex~\cite{Aguilar:2024dlv}, and four- and five-body calculations~\cite{Eichmann:2015cra,Wallbott:2019dng,Hoffer:2024alv,Eichmann:2025gyz}.

\newpage

\section{Conclusions}\label{sec:conclusions}

The structure of $n$-point functions can be intimidating, especially when it comes to four-, five- and six-point functions or above. 
However, there are tools available to simplify their properties.
In Section~\ref{sec:tensors} we saw that orthonormal tensor bases make it easy to count the number of independent Dirac-Lorentz tensors,
and they are very useful for projecting out Lorentz-invariant equations.
In Sections~\ref{sec:tensorbases} and~\ref{sec:li} we highlighted the role of symmetries in organizing the tensor basis elements and Lorentz invariants.
If the $n$-point function is subject to permutation symmetries, these can be implemented  at the level of the basis elements 
so that the dressing function become singlets under permutations. 
As a consequence, the dressing functions often show a planar degeneracy in terms of a minimal angular dependence,
which effectively reduces their kinematic dependence to only a few variables or even just the symmetric variable only.
Moreover, the implementation of gauge symmetries  implies a power counting which helps to identify the leading components.
Taken together, these tools can greatly reduce the numerical effort in dealing with higher $n$-point functions 
and  assist in their better understanding -- often the dynamical content of some amplitude
with many tensors and many Lorentz invariants is stored in just a few dressing functions
which depend on a small number of momentum variables.

To summarize, the bottom line for beginners is: \textit{Don't be afraid of Dirac and Lorentz indices!} 
Symmetries are here to help -- they make your life easier, your code faster, and they sharpen your physical intuition. 
Last but not least, they are fun to play with!

\vspace{6pt}

\funding{This research was funded in whole, or in part, by the Austrian Science Fund (FWF) under grant number 10.55776/PAT2089624. 
For the purpose of open access, the author has applied a CC BY public copyright licence to any Author Accepted Manuscript version arising from this submission.
The author acknowledges the financial support by the University of Graz.}
 
\informedconsent{Not applicable.}

\dataavailability{Data sharing is not applicable to this article.} 

\acknowledgments{I would like to thank Craig Roberts and the organizers of the Workshop and School on Hadron Structure and Strong Interactions (WSHSSI),
Nanjing University, Oct. 13--17, 2025. The contents of this work were originally
prepared for a lecture series given at the ECT$^\ast$ Doctoral Training Program ``Hadron physics with functional methods'', ECT$^\ast$, Trento, Italy,
May 2--18, 2022.}

\conflictsofinterest{The author declares no conflicts of interest. The funders had no role in the design of the study; in the collection, analyses, or interpretation of data; in the writing of the manuscript; or in the decision to publish the results.}

\begin{adjustwidth}{-\extralength}{0cm}

\reftitle{References}



\bibliography{reference}

\begin{thebibliography}{999}

\bibitem[Eichmann(2011)]{Eichmann:2011vu}
Eichmann, G.
\newblock {Nucleon electromagnetic form factors from the covariant Faddeev
  equation}.
\newblock {\em Phys. Rev. D} {\bf 2011}, {\em 84},~014014,
  \href{http://arxiv.org/abs/1104.4505}{{\normalfont
  [arXiv:hep-ph/1104.4505]}}.
\newblock {\url{https://doi.org/10.1103/PhysRevD.84.014014}}.

\bibitem[Eichmann and Fischer(2013)]{Eichmann:2012mp}
Eichmann, G.; Fischer, C.S.
\newblock {Nucleon Compton scattering in the Dyson-Schwinger approach}.
\newblock {\em Phys. Rev. D} {\bf 2013}, {\em 87},~036006,
  \href{http://arxiv.org/abs/1212.1761}{{\normalfont
  [arXiv:hep-ph/1212.1761]}}.
\newblock {\url{https://doi.org/10.1103/PhysRevD.87.036006}}.

\bibitem[Eichmann et~al.(2014)Eichmann, Williams, Alkofer, and
  Vujinovic]{Eichmann:2014xya}
Eichmann, G.; Williams, R.; Alkofer, R.; Vujinovic, M.
\newblock Three-gluon vertex in Landau gauge.
\newblock {\em Phys. Rev. D} {\bf 2014}, {\em 89},~105014,
  \href{http://arxiv.org/abs/1402.1365}{{\normalfont
  [arXiv:hep-ph/1402.1365]}}.
\newblock {\url{https://doi.org/10.1103/PhysRevD.89.105014}}.

\bibitem[Eichmann et~al.(2015)Eichmann, Fischer, and Heupel]{Eichmann:2015nra}
Eichmann, G.; Fischer, C.S.; Heupel, W.
\newblock Four-point functions and the permutation group S4.
\newblock {\em Phys. Rev. D} {\bf 2015}, {\em 92},~056006,
  \href{http://arxiv.org/abs/1505.06336}{{\normalfont
  [arXiv:hep-ph/1505.06336]}}.
\newblock {\url{https://doi.org/10.1103/PhysRevD.92.056006}}.

\bibitem[Eichmann et~al.(2016)Eichmann, Sanchis-Alepuz, Williams, Alkofer, and
  Fischer]{Eichmann:2016yit}
Eichmann, G.; Sanchis-Alepuz, H.; Williams, R.; Alkofer, R.; Fischer, C.S.
\newblock {Baryons as relativistic three-quark bound states}.
\newblock {\em Prog. Part. Nucl. Phys.} {\bf 2016}, {\em 91},~1--100,
  \href{http://arxiv.org/abs/1606.09602}{{\normalfont
  [arXiv:hep-ph/1606.09602]}}.
\newblock {\url{https://doi.org/10.1016/j.ppnp.2016.07.001}}.

\bibitem[Sanchis-Alepuz and Williams(2018)]{Sanchis-Alepuz:2017jjd}
Sanchis-Alepuz, H.; Williams, R.
\newblock Recent developments in bound-state calculations using the
  Dyson–Schwinger and Bethe–Salpeter equations.
\newblock {\em Comput. Phys. Commun.} {\bf 2018}, {\em 232},~1--21,
  \href{http://arxiv.org/abs/1710.04903}{{\normalfont
  [arXiv:hep-ph/1710.04903]}}.
\newblock {\url{https://doi.org/10.1016/j.cpc.2018.05.020}}.

\bibitem[Aguilar et~al.(2023)Aguilar, Ferreira, Papavassiliou, and
  Santos]{Aguilar:2023qqd}
Aguilar, A.C.; Ferreira, M.N.; Papavassiliou, J.; Santos, L.R.
\newblock Planar degeneracy of the three-gluon vertex.
\newblock {\em Eur. Phys. J. C} {\bf 2023}, {\em 83},~549,
  \href{http://arxiv.org/abs/2305.05704}{{\normalfont
  [arXiv:hep-ph/2305.05704]}}.
\newblock {\url{https://doi.org/10.1140/epjc/s10052-023-11732-3}}.

\bibitem[Eichmann and Torres(2025)]{Eichmann:2025etg}
Eichmann, G.; Torres, R.D.
\newblock Five-point functions and the permutation group S5.
\newblock {\em Phys. Rev. D} {\bf 2025}, {\em 111},~094008,
  \href{http://arxiv.org/abs/2502.17225}{{\normalfont
  [arXiv:hep-ph/2502.17225]}}.
\newblock {\url{https://doi.org/10.1103/PhysRevD.111.094008}}.

\bibitem[Braun et~al.(2026)Braun, Gei{\ss}el, Pawlowski, Sattler, and
  Wink]{Braun:2025gvq}
Braun, J.; Gei{\ss}el, A.; Pawlowski, J.M.; Sattler, F.R.; Wink, N.
\newblock Juggling with tensor bases in functional approaches.
\newblock {\em Annals Phys.} {\bf 2026}, {\em 484},~170250,
  \href{http://arxiv.org/abs/2503.05580}{{\normalfont
  [arXiv:hep-th/2503.05580]}}.
\newblock {\url{https://doi.org/10.1016/j.aop.2025.170250}}.

\bibitem[Eichmann(2025)]{Eichmann:2025wgs}
Eichmann, G.
\newblock Hadron physics with functional methods {\bf 2025}.
\newblock  \href{http://arxiv.org/abs/2503.10397}{{\normalfont
  [arXiv:hep-ph/2503.10397]}}.

\bibitem[Huber(2025)]{Huber:2025cbd}
Huber, M.Q.
\newblock A beginner's guide to functional methods in particle physics {\bf
  2025}.
\newblock  \href{http://arxiv.org/abs/2510.18960}{{\normalfont
  [arXiv:hep-ph/2510.18960]}}.

\bibitem[Eichmann and Ramalho(2018)]{Eichmann:2018ytt}
Eichmann, G.; Ramalho, G.
\newblock Nucleon resonances in Compton scattering.
\newblock {\em Phys. Rev. D} {\bf 2018}, {\em 98},~093007,
  \href{http://arxiv.org/abs/1806.04579}{{\normalfont
  [arXiv:hep-ph/1806.04579]}}.
\newblock {\url{https://doi.org/10.1103/PhysRevD.98.093007}}.

\bibitem[Wallbott et~al.(2019)Wallbott, Eichmann, and
  Fischer]{Wallbott:2019dng}
Wallbott, P.C.; Eichmann, G.; Fischer, C.S.
\newblock X(3872) as a four-quark state in a Dyson-Schwinger/Bethe-Salpeter
  approach.
\newblock {\em Phys. Rev. D} {\bf 2019}, {\em 100},~014033,
  \href{http://arxiv.org/abs/1905.02615}{{\normalfont
  [arXiv:hep-ph/1905.02615]}}.
\newblock {\url{https://doi.org/10.1103/PhysRevD.100.014033}}.

\bibitem[Eichmann et~al.(2016)Eichmann, Fischer, Heupel, and
  Williams]{Eichmann:2014ooa}
Eichmann, G.; Fischer, C.S.; Heupel, W.; Williams, R.
\newblock The muon g-2: Dyson-Schwinger status on hadronic light-by-light
  scattering.
\newblock {\em AIP Conf. Proc.} {\bf 2016}, {\em 1701},~040004,
  \href{http://arxiv.org/abs/1411.7876}{{\normalfont
  [arXiv:hep-ph/1411.7876]}}.
\newblock {\url{https://doi.org/10.1063/1.4938621}}.

\bibitem[Aguilar et~al.(2012)Aguilar, Ibanez, Mathieu, and
  Papavassiliou]{Aguilar:2011xe}
Aguilar, A.C.; Ibanez, D.; Mathieu, V.; Papavassiliou, J.
\newblock Massless bound-state excitations and the Schwinger mechanism in QCD.
\newblock {\em Phys. Rev. D} {\bf 2012}, {\em 85},~014018,
  \href{http://arxiv.org/abs/1110.2633}{{\normalfont
  [arXiv:hep-ph/1110.2633]}}.
\newblock {\url{https://doi.org/10.1103/PhysRevD.85.014018}}.

\bibitem[Aguilar et~al.(2017)Aguilar, Binosi, and
  Papavassiliou]{Aguilar:2016ock}
Aguilar, A.C.; Binosi, D.; Papavassiliou, J.
\newblock Schwinger mechanism in linear covariant gauges.
\newblock {\em Phys. Rev. D} {\bf 2017}, {\em 95},~034017,
  \href{http://arxiv.org/abs/1611.02096}{{\normalfont
  [arXiv:hep-ph/1611.02096]}}.
\newblock {\url{https://doi.org/10.1103/PhysRevD.95.034017}}.

\bibitem[Eichmann et~al.(2021)Eichmann, Pawlowski, and Silva]{Eichmann:2021zuv}
Eichmann, G.; Pawlowski, J.M.; Silva, J.M.
\newblock {Mass generation in Landau-gauge Yang-Mills theory}.
\newblock {\em Phys. Rev. D} {\bf 2021}, {\em 104},~114016,
  \href{http://arxiv.org/abs/2107.05352}{{\normalfont
  [arXiv:hep-ph/2107.05352]}}.
\newblock {\url{https://doi.org/10.1103/PhysRevD.104.114016}}.

\bibitem[Aguilar et~al.(2022)Aguilar, Ferreira, and
  Papavassiliou]{Aguilar:2021uwa}
Aguilar, A.C.; Ferreira, M.N.; Papavassiliou, J.
\newblock Exploring smoking-gun signals of the Schwinger mechanism in QCD.
\newblock {\em Phys. Rev. D} {\bf 2022}, {\em 105},~014030,
  \href{http://arxiv.org/abs/2111.09431}{{\normalfont
  [arXiv:hep-ph/2111.09431]}}.
\newblock {\url{https://doi.org/10.1103/PhysRevD.105.014030}}.

\bibitem[Aguilar et~al.(2023)Aguilar, De~Soto, Ferreira, Papavassiliou,
  Pinto-G{\'o}mez, Roberts, and Rodr{\'\i}guez-Quintero]{Aguilar:2022thg}
Aguilar, A.C.; De~Soto, F.; Ferreira, M.N.; Papavassiliou, J.; Pinto-G{\'o}mez,
  F.; Roberts, C.D.; Rodr{\'\i}guez-Quintero, J.
\newblock Schwinger mechanism for gluons from lattice QCD.
\newblock {\em Phys. Lett. B} {\bf 2023}, {\em 841},~137906,
  \href{http://arxiv.org/abs/2211.12594}{{\normalfont
  [arXiv:hep-ph/2211.12594]}}.
\newblock {\url{https://doi.org/10.1016/j.physletb.2023.137906}}.

\bibitem[Ball and Chiu(1980)]{Ball:1980ax}
Ball, J.S.; Chiu, T.W.
\newblock Analytic Properties of the Vertex Function in Gauge Theories. 2.
\newblock {\em Phys. Rev. D} {\bf 1980}, {\em 22},~2550.
\newblock [Erratum: Phys.Rev.D 23, 3085 (1981)],
  {\url{https://doi.org/10.1103/PhysRevD.22.2550}}.

\bibitem[Kizilersu et~al.(1995)Kizilersu, Reenders, and
  Pennington]{Kizilersu:1995iz}
Kizilersu, A.; Reenders, M.; Pennington, M.R.
\newblock One loop QED vertex in any covariant gauge: Its complete analytic
  form.
\newblock {\em Phys. Rev. D} {\bf 1995}, {\em 52},~1242--1259,
  \href{http://arxiv.org/abs/hep-ph/9503238}{{\normalfont [hep-ph/9503238]}}.
\newblock {\url{https://doi.org/10.1103/PhysRevD.52.1242}}.

\bibitem[Davydychev et~al.(2001)Davydychev, Osland, and
  Saks]{Davydychev:2000rt}
Davydychev, A.I.; Osland, P.; Saks, L.
\newblock Quark gluon vertex in arbitrary gauge and dimension.
\newblock {\em Phys. Rev. D} {\bf 2001}, {\em 63},~014022,
  \href{http://arxiv.org/abs/hep-ph/0008171}{{\normalfont [hep-ph/0008171]}}.
\newblock {\url{https://doi.org/10.1103/PhysRevD.63.014022}}.

\bibitem[Skullerud and Kizilersu(2002)]{Skullerud:2002ge}
Skullerud, J.; Kizilersu, A.
\newblock Quark gluon vertex from lattice QCD.
\newblock {\em JHEP} {\bf 2002}, {\em 09},~013,
  \href{http://arxiv.org/abs/hep-ph/0205318}{{\normalfont [hep-ph/0205318]}}.
\newblock {\url{https://doi.org/10.1088/1126-6708/2002/09/013}}.

\bibitem[Bashir et~al.(2012)Bashir, Bermudez, Chang, and
  Roberts]{Bashir:2011dp}
Bashir, A.; Bermudez, R.; Chang, L.; Roberts, C.D.
\newblock Dynamical chiral symmetry breaking and the fermion–gauge-boson
  vertex.
\newblock {\em Phys. Rev. C} {\bf 2012}, {\em 85},~045205,
  \href{http://arxiv.org/abs/1112.4847}{{\normalfont
  [arXiv:nucl-th/1112.4847]}}.
\newblock {\url{https://doi.org/10.1103/PhysRevC.85.045205}}.

\bibitem[Maris and Tandy(2000)]{Maris:1999bh}
Maris, P.; Tandy, P.C.
\newblock {The Quark photon vertex and the pion charge radius}.
\newblock {\em Phys. Rev. C} {\bf 2000}, {\em 61},~045202,
  \href{http://arxiv.org/abs/nucl-th/9910033}{{\normalfont [nucl-th/9910033]}}.
\newblock {\url{https://doi.org/10.1103/PhysRevC.61.045202}}.

\bibitem[Maris and Tandy(2006)]{Maris:2005tt}
Maris, P.; Tandy, P.C.
\newblock {QCD modeling of hadron physics}.
\newblock {\em Nucl. Phys. B Proc. Suppl.} {\bf 2006}, {\em 161},~136--152,
  \href{http://arxiv.org/abs/nucl-th/0511017}{{\normalfont [nucl-th/0511017]}}.
\newblock {\url{https://doi.org/10.1016/j.nuclphysbps.2006.08.012}}.

\bibitem[Krassnigg and Maris(2005)]{Krassnigg:2004if}
Krassnigg, A.; Maris, P.
\newblock Pseudoscalar and vector mesons as q anti-q bound states.
\newblock {\em J. Phys. Conf. Ser.} {\bf 2005}, {\em 9},~153--160,
  \href{http://arxiv.org/abs/nucl-th/0412058}{{\normalfont [nucl-th/0412058]}}.
\newblock {\url{https://doi.org/10.1088/1742-6596/9/1/029}}.

\bibitem[Eichmann et~al.(2020)Eichmann, Fischer, and
  Williams]{Eichmann:2019bqf}
Eichmann, G.; Fischer, C.S.; Williams, R.
\newblock Kaon-box contribution to the anomalous magnetic moment of the muon.
\newblock {\em Phys. Rev. D} {\bf 2020}, {\em 101},~054015,
  \href{http://arxiv.org/abs/1910.06795}{{\normalfont
  [arXiv:hep-ph/1910.06795]}}.
\newblock {\url{https://doi.org/10.1103/PhysRevD.101.054015}}.

\bibitem[Miramontes et~al.(2021)Miramontes, Sanchis~Alepuz, and
  Alkofer]{Miramontes:2021xgn}
Miramontes, A.S.; Sanchis~Alepuz, H.; Alkofer, R.
\newblock {Elucidating the effect of intermediate resonances in the quark
  interaction kernel on the timelike electromagnetic pion form factor}.
\newblock {\em Phys. Rev. D} {\bf 2021}, {\em 103},~116006,
  \href{http://arxiv.org/abs/2102.12541}{{\normalfont
  [arXiv:hep-ph/2102.12541]}}.
\newblock {\url{https://doi.org/10.1103/PhysRevD.103.116006}}.

\bibitem[Eichmann(2014)]{Eichmann:2014qva}
Eichmann, G.
\newblock Probing nucleons with photons at the quark level.
\newblock {\em Acta Phys. Polon. Supp.} {\bf 2014}, {\em 7},~597,
  \href{http://arxiv.org/abs/1404.4149}{{\normalfont
  [arXiv:nucl-th/1404.4149]}}.
\newblock {\url{https://doi.org/10.5506/APhysPolBSupp.7.597}}.

\bibitem[Leutnant and Sternbeck(2018)]{Leutnant:2018dry}
Leutnant, M.; Sternbeck, A.
\newblock Quark-photon vertex from lattice QCD in Landau gauge.
\newblock {\em PoS} {\bf 2018}, {\em Confinement2018},~095,
  \href{http://arxiv.org/abs/1812.11131}{{\normalfont
  [arXiv:hep-lat/1812.11131]}}.
\newblock {\url{https://doi.org/10.22323/1.336.0095}}.

\bibitem[Bardeen and Tung(1968)]{Bardeen:1968ebo}
Bardeen, W.A.; Tung, W.K.
\newblock Invariant amplitudes for photon processes.
\newblock {\em Phys. Rev.} {\bf 1968}, {\em 173},~1423--1433.
\newblock [Erratum: Phys.Rev.D 4, 3229--3229 (1971)],
  {\url{https://doi.org/10.1103/PhysRev.173.1423}}.

\bibitem[Tarrach(1975)]{Tarrach:1975tu}
Tarrach, R.
\newblock Invariant Amplitudes for Virtual Compton Scattering Off Polarized
  Nucleons Free from Kinematical Singularities, Zeros and Constraints.
\newblock {\em Nuovo Cim. A} {\bf 1975}, {\em 28},~409.
\newblock {\url{https://doi.org/10.1007/BF02894857}}.

\bibitem[Drechsel et~al.(2003)Drechsel, Pasquini, and
  Vanderhaeghen]{Drechsel:2002ar}
Drechsel, D.; Pasquini, B.; Vanderhaeghen, M.
\newblock Dispersion relations in real and virtual Compton scattering.
\newblock {\em Phys. Rept.} {\bf 2003}, {\em 378},~99--205,
  \href{http://arxiv.org/abs/hep-ph/0212124}{{\normalfont [hep-ph/0212124]}}.
\newblock {\url{https://doi.org/10.1016/S0370-1573(02)00636-1}}.

\bibitem[Colangelo et~al.(2015)Colangelo, Hoferichter, Procura, and
  Stoffer]{Colangelo:2015ama}
Colangelo, G.; Hoferichter, M.; Procura, M.; Stoffer, P.
\newblock Dispersion relation for hadronic light-by-light scattering:
  theoretical foundations.
\newblock {\em JHEP} {\bf 2015}, {\em 09},~074,
  \href{http://arxiv.org/abs/1506.01386}{{\normalfont
  [arXiv:hep-ph/1506.01386]}}.
\newblock {\url{https://doi.org/10.1007/JHEP09(2015)074}}.

\bibitem[Colangelo et~al.(2017)Colangelo, Hoferichter, Procura, and
  Stoffer]{Colangelo:2017fiz}
Colangelo, G.; Hoferichter, M.; Procura, M.; Stoffer, P.
\newblock Dispersion relation for hadronic light-by-light scattering: two-pion
  contributions.
\newblock {\em JHEP} {\bf 2017}, {\em 04},~161,
  \href{http://arxiv.org/abs/1702.07347}{{\normalfont
  [arXiv:hep-ph/1702.07347]}}.
\newblock {\url{https://doi.org/10.1007/JHEP04(2017)161}}.

\bibitem[Miramontes et~al.(2025)Miramontes, Eichmann, and
  Alkofer]{Miramontes:2025ofw}
Miramontes, A.S.; Eichmann, G.; Alkofer, R.
\newblock Timelike form factor for the anomalous process $\gamma^\ast \pi \to
  \pi \pi$.
\newblock {\em Phys. Lett. B} {\bf 2025}, {\em 868},~139659,
  \href{http://arxiv.org/abs/2504.20899}{{\normalfont
  [arXiv:hep-ph/2504.20899]}}.
\newblock {\url{https://doi.org/10.1016/j.physletb.2025.139659}}.

\bibitem[Aguilar et~al.(2024)Aguilar, De~Soto, Ferreira, Papavassiliou,
  Pinto-G\'omez, Rodr\'\i{}guez-Quintero, and Santos]{Aguilar:2024dlv}
Aguilar, A.C.; De~Soto, F.; Ferreira, M.N.; Papavassiliou, J.; Pinto-G\'omez,
  F.; Rodr\'\i{}guez-Quintero, J.; Santos, L.R.
\newblock {Nonperturbative four-gluon vertex in soft kinematics}.
\newblock {\em Phys. Lett. B} {\bf 2024}, {\em 858},~139065,
  \href{http://arxiv.org/abs/2408.06135}{{\normalfont
  [arXiv:hep-ph/2408.06135]}}.
\newblock {\url{https://doi.org/10.1016/j.physletb.2024.139065}}.

\bibitem[Eichmann et~al.(2016)Eichmann, Fischer, and Heupel]{Eichmann:2015cra}
Eichmann, G.; Fischer, C.S.; Heupel, W.
\newblock {The light scalar mesons as tetraquarks}.
\newblock {\em Phys. Lett. B} {\bf 2016}, {\em 753},~282--287,
  \href{http://arxiv.org/abs/1508.07178}{{\normalfont
  [arXiv:hep-ph/1508.07178]}}.
\newblock {\url{https://doi.org/10.1016/j.physletb.2015.12.036}}.

\bibitem[Hoffer et~al.(2024)Hoffer, Eichmann, and Fischer]{Hoffer:2024alv}
Hoffer, J.; Eichmann, G.; Fischer, C.S.
\newblock {Hidden-flavor four-quark states in the charm and bottom region}.
\newblock {\em Phys. Rev. D} {\bf 2024}, {\em 109},~074025,
  \href{http://arxiv.org/abs/2402.12830}{{\normalfont
  [arXiv:hep-ph/2402.12830]}}.
\newblock {\url{https://doi.org/10.1103/PhysRevD.109.074025}}.

\bibitem[Eichmann et~al.(2025)Eichmann, Pe{\~n}a, and Torres]{Eichmann:2025gyz}
Eichmann, G.; Pe{\~n}a, M.T.; Torres, R.D.
\newblock Five-body systems with Bethe-Salpeter equations.
\newblock {\em Phys. Lett. B} {\bf 2025}, {\em 866},~139525,
  \href{http://arxiv.org/abs/2502.17944}{{\normalfont
  [arXiv:hep-ph/2502.17944]}}.
\newblock {\url{https://doi.org/10.1016/j.physletb.2025.139525}}.

\end{thebibliography}

\PublishersNote{}
\end{adjustwidth}
\end{document}